\documentclass[11pt,a4paper]{article}
\pdfoutput=1
\usepackage{jinstpub}
\usepackage{graphicx}% Include figure files amsmath,amssymb,aps,floatfix]{revtex4-1}
\usepackage{epstopdf}
\usepackage{color}
\usepackage{dcolumn}% Align table columns on decimal point
\usepackage{bm}% bold math
\usepackage{hyperref}% add hypertext capabilities
\usepackage{siunitx}
\usepackage{booktabs}
\usepackage[normalem]{ulem} %STRIKE
\usepackage{threeparttable} % for more complex tables
\usepackage{tabularx}
\usepackage{siunitx}
\usepackage{adjustbox}

\newcommand{\isot}[2]{$^{\textrm{#2}}$#1 }
\newcommand{\krm}{\isot{Kr}{83m}}
\newcommand{\subStwo}{$_{\textrm{S2}}$}

\begin{document}
\title{\boldmath 3D modeling of electric fields in the LUX detector}

\author{The LUX collaboration:}
\author[a,b,c]{D.S.~Akerib,}
\author[d]{S.~Alsum,}  
\author[e]{H.M.~Ara\'{u}jo,} 
\author[f]{X.~Bai,}  
\author[e]{A.J.~Bailey,} 
\author[g]{J.~Balajthy,}  
\author[h]{P.~Beltrame,} 
\author[i,j]{E.P.~Bernard,} 
\author[k]{A.~Bernstein,}   
\author[a,b,c]{T.P.~Biesiadzinski,} 
\author[i,j]{E.M.~Boulton,}
\author[l]{P.~Br\'as,}  
\author[m,n]{D.~Byram,} 
\author[j]{S.B.~Cahn,} 
\author[o,p]{M.C.~Carmona-Benitez,} 
\author[q]{C.~Chan,}  
\author[e]{A.~Currie,} 
\author[r]{J.E.~Cutter,}   
\author[h]{T.J.R.~Davison,} 
\author[s]{A.~Dobi,} 
\author[t]{E.~Druszkiewicz,} 
\author[j]{B.N.~Edwards,} 
\author[u]{S.R.~Fallon,}  
\author[b,c]{A.~Fan,}
\author[q,s]{S.~Fiorucci,} 
\author[q]{R.J.~Gaitskell,}  
\author[u]{J.~Genovesi,} 
\author[v]{C.~Ghag,}  
\author[s]{M.G.D.~Gilchriese,}
\author[g]{C.R.~Hall,}   
\author[f,n]{M.~Hanhardt,} 
\author[p]{S.J.~Haselschwardt,}
\author[j,s,w]{S.A.~Hertel,} 
\author[i]{D.P.~Hogan,}
\author[i,j,n]{M.~Horn,} 
\author[q]{D.Q.~Huang,} 
\author[b,c]{C.M.~Ignarra,} 
\author[i]{R.G.~Jacobsen,} 
\author[a,b,c]{W.~Ji,} 
\author[i]{K.~Kamdin,}
\author[k]{K.~Kazkaz,} 
\author[t]{D.~Khaitan,} 
\author[g]{R.~Knoche,} 
\author[j]{N.A.~Larsen,} 
\author[k,r]{B.G.~Lenardo,} 
\author[s]{K.T.~Lesko,} 
\author[l]{A.~Lindote,} 
\author[l]{M.I.~Lopes,} 
\author[r]{A.~Manalaysay,}  
\author[d,x]{R.L.~Mannino,}  
\author[h]{M.F.~Marzioni,}   
\author[i,j,s]{D.N.~McKinsey,} 
\author[m]{D.-M.~Mei,} 
\author[u]{J.~Mock,}
\author[t]{M.~Moongweluwan,} 
\author[r]{J.A.~Morad,} 
\author[h]{A.St.J.~Murphy,} 
\author[p]{C.~Nehrkorn,}
\author[p]{H.N.~Nelson,} 
\author[l]{F.~Neves,}   
\author[i,j,s]{K.~O'Sullivan,}
\author[i]{K.C.~Oliver-Mallory,}
\author[b,c,d]{K.J.~Palladino,} 
\author[i,j]{E.K.~Pease,} 
\author[q]{C.~Rhyne,}
\author[p,v]{S.~Shaw,} 
\author[a,c]{T.A.~Shutt,} 
\author[l]{C.~Silva,}   
\author[p]{M.~Solmaz,}  
\author[l]{V.N.~Solovov,} 
\author[s]{P.~Sorensen,} 
\author[e]{T.J.~Sumner,} 
\author[u]{M.~Szydagis,}   
\author[n]{D.J.~Taylor,} 
\author[q]{W.C.~Taylor,} 
\author[j]{B.P.~Tennyson,} 
\author[x]{P.A.~Terman,} 
\author[f]{D.R.~Tiedt,}  
\author[b,c,y]{W.H.~To,} 
\author[r]{M.~Tripathi,} 
\author[i,j,s,1]{L.~Tvrznikova,\note{Corresponding author}}
\author[r]{S.~Uvarov,}   
\author[i]{V.~Velan,} 
\author[q]{J.R.~Verbus,} 
\author[x]{R.C.~Webb,} 
\author[x]{J.T.~White,}
\author[a,b,c]{T.J.~Whitis,} 
\author[s]{M.S.~Witherell,} 
\author[t]{F.L.H.~Wolfs,}  
\author[k]{J.~Xu,}
\author[e]{K.~Yazdani,}
\author[u]{S.K.~Young,} 
\author[m]{C.~Zhang.} 

\affiliation[a]{Case Western Reserve University, Department of Physics, 10900 Euclid Ave, Cleveland, OH 44106, USA} 
\affiliation[b]{SLAC National Accelerator Laboratory, 2575 Sand Hill Road, Menlo Park, CA 94205, USA} 
\affiliation[c]{Kavli Institute for Particle Astrophysics and Cosmology, Stanford University, 452 Lomita Mall, Stanford, CA 94309, USA}
\affiliation[d]{University of Wisconsin-Madison, Department of Physics, 1150 University Ave., Madison, WI 53706, USA}
\affiliation[e]{Imperial College London, High Energy Physics, Blackett Laboratory, London SW7 2BZ, United Kingdom}  
\affiliation[f]{South Dakota School of Mines and Technology, 501 East St Joseph St., Rapid City, SD 57701, USA} 
\affiliation{Imperial College London, High Energy Physics, Blackett Laboratory, London SW7 2BZ, United Kingdom}  
\affiliation[g]{University of Maryland, Department of Physics, College Park, MD 20742, USA} 
\affiliation[h]{SUPA, School of Physics and Astronomy, University of Edinburgh, Edinburgh EH9 3FD, United Kingdom}  
\affiliation[i]{University of California Berkeley, Department of Physics, Berkeley, CA 94720, USA} 
\affiliation[j]{Yale University, Department of Physics, 217 Prospect St., New Haven, CT 06511, USA} 
\affiliation[k]{Lawrence Livermore National Laboratory, 7000 East Ave., Livermore, CA 94551, USA}
\affiliation[l]{LIP-Coimbra, Department of Physics, University of Coimbra, Rua Larga, 3004-516 Coimbra, Portugal} 
 \affiliation[m]{University of South Dakota, Department of Physics, 414E Clark St., Vermillion, SD 57069, USA} 
\affiliation[n]{South Dakota Science and Technology Authority, Sanford Underground Research Facility, Lead, SD 57754, USA}
\affiliation[o]{Pennsylvania State University, Department of Physics, 104 Davey Lab, University Park, PA  16802-6300, USA} 
\affiliation[p]{University of California Santa Barbara, Department of Physics, Santa Barbara, CA 93106, USA} 
\affiliation[q]{Brown University, Department of Physics, 182 Hope St., Providence, RI 02912, USA} 
\affiliation[r]{University of California Davis, Department of Physics, One Shields Ave., Davis, CA 95616, USA}
\affiliation[s]{Lawrence Berkeley National Laboratory, 1 Cyclotron Rd., Berkeley, CA 94720, USA}  
\affiliation[t]{University of Rochester, Department of Physics and Astronomy, Rochester, NY 14627, USA}  
\affiliation[u]{University at Albany, State University of New York, Department of Physics, 1400 Washington Ave., Albany, NY 12222, USA} 
\affiliation[v]{Department of Physics and Astronomy, University College London, Gower Street, London WC1E 6BT, United Kingdom} 
\affiliation[w]{University of Massachusetts, Amherst Center for Fundamental Interactions and Department of Physics, Amherst, MA 01003-9337 USA} 
\affiliation[x]{Texas A \& M University, Department of Physics, College Station, TX 77843, USA}
\affiliation[y]{California State University Stanislaus, Department of Physics, 1 University Circle, Turlock, CA 95382, USA} 

%\collaboration{LUX Collaboration}
\emailAdd{lucie.tvrznikova@yale.edu}

\abstract{
This work details the development of a three-dimensional (3D) electric field model for the LUX detector. The detector took data to search for weakly interacting massive particles (WIMPs) during two periods. After the first period completed, a time-varying non-uniform negative charge developed in the polytetrafluoroethylene (PTFE) panels that define the radial boundary of the detector's active volume. This caused electric field variations in the detector in time, depth and azimuth, generating an electrostatic radially-inward force on electrons on their way upward to the liquid surface. To map this behavior, 3D electric field maps of the detector's active volume were generated on a monthly basis. This was done by fitting a model built in \textsc{COMSOL} Multiphysics to the uniformly distributed calibration data that were collected on a regular basis. The modeled average PTFE charge density increased over the course of the exposure from -3.6 to $-5.5~\mu$C/m$^2$. From our studies, we deduce that the electric field magnitude varied locally while the mean value of the field of $\sim200$~V/cm remained constant throughout the exposure. As a result of this work the varying electric fields and their impact on event reconstruction and discrimination were successfully modeled.
}

\arxivnumber{1709.00095}

\maketitle
\flushbottom

\section{The LUX detector}
The Large Underground Xenon (LUX) experiment is a two-phase (liquid-gas) xenon time projection chamber (TPC) searching for weakly interacting massive particles, a favored dark matter candidate. The detector uses a dodecagonal prism shaped active volume bounded in \textit{z} by cathode and gate wire grids and in (\textit{x, y}) by 12 polytetrafluoroethylene (PTFE) panels and contains 251~kg of ultra pure xenon. The inner cryostat is 101~cm tall and 62~cm in diameter. The active volume diameter, as measured at -100$^{\circ}$C between parallel opposite faces, is 47.3~cm. The distance between cathode and gate grids is 48.3~cm and distance between gate and anode grids is 1~cm. 

Energy deposited by particle interactions induces two measurable signals in the active volume: a prompt scintillation of vacuum ultraviolet (VUV) photons (S1) and an ionization signal from the liberated electrons (S2). The S1 photons are emitted at the interaction site and are detected by top and bottom arrays of 61 photomultiplier tubes (PMTs) each. The detector's drift field pulls free electrons to the liquid surface where they are extracted into the gas, producing a secondary electroluminescence S2 signal on their path to the anode. The S2 photons are detected by both PMT arrays with localization of the signal in the top array. The PMT signals from both light pulses, S1 and S2, along with their time separation that defines the depth of the interaction, allow the reconstruction of interaction vertices in three dimensions (\textit{x\subStwo, y\subStwo, t}). The ratio of free charge to scintillation light, typically expressed as $\log_{10}$(S2/S1), is dependent on the electric field at the event site and is used to distinguish electronic recoils (ERs) and nuclear recoil (NRs) produced by incoming particles interacting with xenon atoms. The NR signals exhibit less variation as a function of field in comparison to ERs, due to the lower ionization of NRs and thus lower importance of electron-ion recombination. This enables discrimination between ER and NR events and makes liquid xenon TPC detectors viable dark matter discovery experiments. Leveraging these capabilities helped LUX produce world-leading exclusion limits for both spin-independent and spin-dependent WIMP-nucleon scattering cross-sections, axion searches, and pioneer many innovative calibration techniques~\cite{Akerib:2013tjd,Akerib:2015rjg,Akerib:2016lao,Akerib:2016vxi,Akerib:2017uem,Akerib:2017kat}. Further technical details about the LUX detector can be found in~\cite{Akerib:2012ys}.

\begin{figure}[ht!]
\begin{center}
\includegraphics[width=0.4\textwidth,clip]{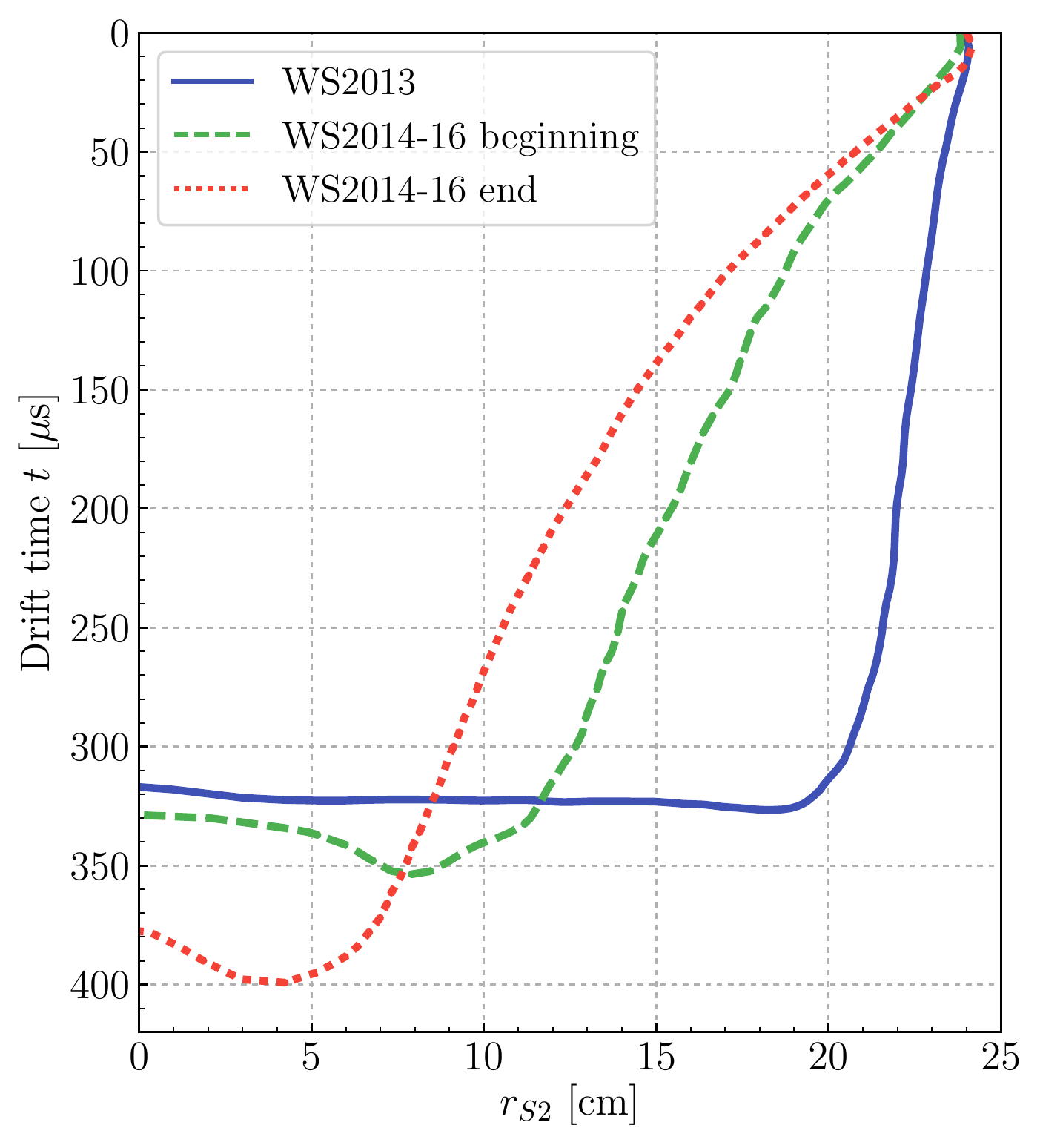}
\caption{Contour plot illustrating detectors edges from the reconstructed positions of the \krm calibration data. Differences between WS2013 (data from 2013-05-10, solid blue) and the time-dependent WS2014-16 (data from 2014-09-03, dashed green and 2016-05-03, dotted red) are clearly seen. Due to the changing electric fields, \krm calibration data from the later science run had events with greater drift times.}
\label{fig:contour_runs34}
\end{center} 
\end{figure}

The detector took WIMP search data during two periods; from April 21 to August 8, 2013 (WS2013) and from September 11, 2014 until May 2, 2016 (WS2014-16). Between WS2013 and WS2014-16, from January to March 2014, the detector grids were conditioned to improve the voltages at which the electrodes could be biased. Afterward, the detector was kept at room temperature for one month in April 2014 before starting WS2014-16. During the conditioning, potentials were held just above the onset of sustained discharge and maintained for many days, akin to the burn-in period often employed in room-temperature proportional counter commissioning~\cite{LHCb,ARNISON1990431,DEWULF1988109,AREFIEV198971}. First, the potentials on gate and anode grids were raised to $-5$~kV and 7~kV, respectively, while in cold xenon gas (180~K). Subsequently, the potential on the cathode was raised to -20~kV while in xenon gas at room temperature (296~K). As a result of conditioning, a higher applied anode voltage enabled increased electron extraction efficiency (i.e. the fraction of electrons which promptly cross the liquid-gas interface) from $49\pm3\%$ in WS2013 to $73\pm4\%$ in WS2014-16~\cite{Akerib:2016vxi}. After the conditioning campaign, \krm calibration data (see Section \ref{EFields_Modeling}) revealed that the electron drift trajectories were significantly altered from the regular near-vertical paths observed in WS2013. figure~\ref{fig:contour_runs34} shows the reconstructed radius and drift time of \krm calibration events next to the PTFE wall, before and after conditioning. A strong bias in azimuthal angle and $z$ was observed, and seen to change throughout WS2014-16. The contours were created by finding the average number of events in all the non-empty bins of a 2D histogram in ($r^2$\subStwo, $t$) and then drawing a contour through those bins that contained half of the average number of events. Therefore the contours approximate the edges of the detector's active volume.

We attribute the time evolution of the electric field to the migration of negative charges (electrons) and positive charges (holes) in the PTFE panels that define the radial boundary of the active volume and serve as a reflector to VUV scintillation photons~\cite{coimbra}. Discharges in the xenon gas during conditioning created VUV photons with energies greater than the band gap of PTFE~\cite{KANIK1996455,Policarpo,ANDRESEN1977371,Munsoo,seki1990electronic}, which in turn created electron-hole pairs within $1-2~\mu$m of the PTFE panels~\cite{Zhang20061995}. The behavior of charges in PTFE and other Teflon materials has been widely studied at room and elevated temperatures~\cite{paulmier2014radiation,green2006ptferho,mellinger2004charge,zhang1991constant}, but not at liquid xenon (LXe) temperatures. The energy levels of the electron and hole traps have been measured to be $0.85-1.0$~eV and $0.80-0.90$~eV respectively, showing a slightly higher energy trap for electrons~\cite{Zhang20061995}. PTFE is known to be a strong insulator with dark-current resistivity of $\rho\sim3\times10^{20}~\Omega\cdot\mathrm{cm}$ measured at $5\times10^{-6}~\mathrm{torr}$ at $\ang{25}\mathrm{C}$~\cite{green2006ptferho} and is able to maintain negative charge densities of at least 0.1~mC/m$^2$~\cite{kressmann1996electrets}. This results in a faster removal of holes compared to electrons under the applied field creating the observed abundance of negative charges in the PTFE panels. Furthermore, this behavior varies over time thus introducing a strong radially-inward drift of electrons during WS2014-16~\cite{Akerib:2016vxi} varying in time, depth and azimuth as described in section~\ref{sub:charge_evolution}.

Knowledge of the electric field inside the detector is necessary for (\textit{x, y}) position reconstruction since the drift path of the S2 electrons is field-dependent. Event reconstruction is also field-dependent due to the effect of the drift field on recombination. Since the conditioning altered the electric fields inside the detector in a non-symmetric manner, a full 3D model of the fields was needed throughout the 23~month-long WS2014-16. This paper describes the method used to develop these field models.

\section{Modeling electric fields inside the LUX detector}\label{Modeling_Efields}

Electric field models were developed both with and without charge densities in the PTFE panels to understand the detector behavior. Prior to developing a model for the changing electric fields during WS2014-16, a 3D model was developed without static charges in the PTFE panels to confirm our understanding of the calibration data from WS2013. The 3D model was built using the AC/DC Module of \textsc{COMSOL} Multiphysics v5.0\textsuperscript{\textregistered}~\cite{comsolRef}, a commercially available finite element simulation software.

\subsection{LUX 3D model geometry}
\begin{figure}[ht!]
\begin{center}
\includegraphics[width=0.6\textwidth,clip]{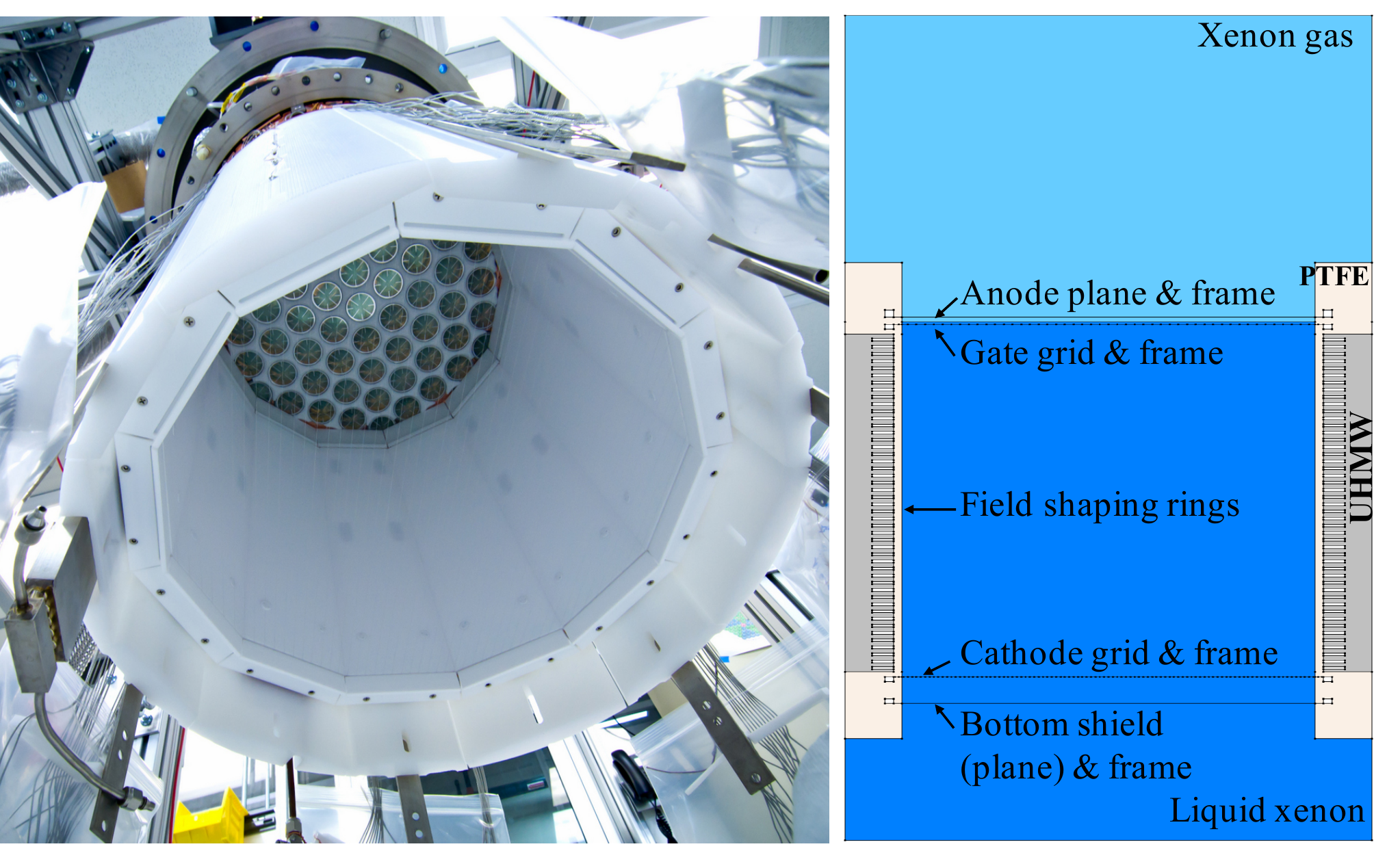}
\caption{Left: bottom view of the LUX detector showing the regular dodecagonal prism shape, the PTFE panels, and the top PMT array. Right: cross-section of the 3D LUX \textsc{COMSOL} model.}
\label{fig:lux_geometry}
\end{center} 
\end{figure}

A full 3D model was needed due to the detector's geometry. The active volume takes a shape of a regular dodecagonal prism as shown in the left panel of figure~\ref{fig:lux_geometry} and the five detector grids (bottom shield, cathode, gate, anode, top shield) are rotated by 60$^{\circ}$~with respect to each other, leaving a geometry that is not azimuthally or otherwise symmetric. The 2D cross-section of the model used in the 3D simulations is shown in the right panel of figure~\ref{fig:lux_geometry}; the figure also illustrates some of the adopted simplifications. Due to the high degree of complexity of the detector, only the active region of the detector was modeled in detail; the bottom shield and anode grids were modeled as planes imposing electrostatic boundaries for the height of the active region of the detector.

Since the model spans 5 orders of magnitude in length scale (from $28~\mu$m for the grid wire diameter to 1.1~m for the inner vacuum cryostat height), the cathode and gate grid wires were modeled as parallel lines with zero diameter to accelerate the convergence of the model. As with all simplifications, test cases were studied to ensure this wire diameter change had a negligible effect on the resulting solution without a loss of field resolution~\cite{mcdonald2003notes}. Further details about the detector grids along with applied grid voltages from WS2013 and WS2014-16 are summarized in Table~\ref{Table:grid_properties}. The details of the ultra-high-molecular-weight polyethylene (UHMW) and PTFE structures surrounding the active volume were simplified and large metal objects residing outside of the field cage (i.e. cathode high voltage cable, heat exchanger) were omitted since a simulation confirmed they do not affect the electric field within the active volume. The relative DC dielectric constants of the materials used in the model were: $\varepsilon_{liquid~xenon}=1.95$, $\varepsilon_{gaseous~xenon}=1.0$, $\varepsilon_{PTFE}=2.1$, $\varepsilon_{UHMW}=2.3$.

\begin{table}[ht]
\setlength{\extrarowheight}{2pt}
\caption{Grid properties and voltages as relevant to the construction of the electric field model used in WS2013 and WS2014-16 simulations, including description of geometric simplifications.
\label{Table:grid_properties}}
%\centering{}%
\begin{center}
%\scalebox{0.95}
%{
\begin{threeparttable}
%\begin{tabular}{>{\raggedright}m{1.9cm} m{1.1cm} m{1.3cm} m{1.2cm} m{1.2cm} m{0.95cm}}
\begin{tabular}{|l *{5}{S[table-format=-3.2]}|}
\hline
{Grid} & {Bottom} & {Cathode} & {Gate} & {Anode} & {Top} \\
{} & {shield} & {} & {} & {} & {shield} \\
\hline
$z$\tnote{\dag} ~[cm] & 2.0 & 5.6 & 53.9 & 54.9 & 58.6 \\
Wire $\varnothing$~[$\mu$m] & 206.0 & 206.0 & 101.6 & 28.4 & 50.8 \\
Pitch [mm] & 10.00 & 5.00 & 5.00 & 0.25 & 5.00 \\
Angle [deg] & 15 & 75 & 15 & N/A & 135 \\
Modeled as &  {Plane}  & {$\varnothing$0~wires}  & {$\varnothing$0~wires} & {Plane} & {Absent} \\
WS2013 [kV] & -2 & -10 & -1.5 & 3.5 & -1 \\
WS2014-16 [kV] & -2 & -8.5 & 1 & 7 & -1 \\
\hline
\end{tabular}
\begin{tablenotes}
\item[\dag] $z$ is defined as vertical distance from the face of the bottom PMT array, accounting for thermal contraction as appropriate.
\end{tablenotes}
\end{threeparttable}
%}
\end{center}
\end{table}

\subsection{Electric fields modeling}\label{EFields_Modeling}

\begin{figure}[ht!]
\begin{center}
\includegraphics[width=0.7\textwidth,clip]{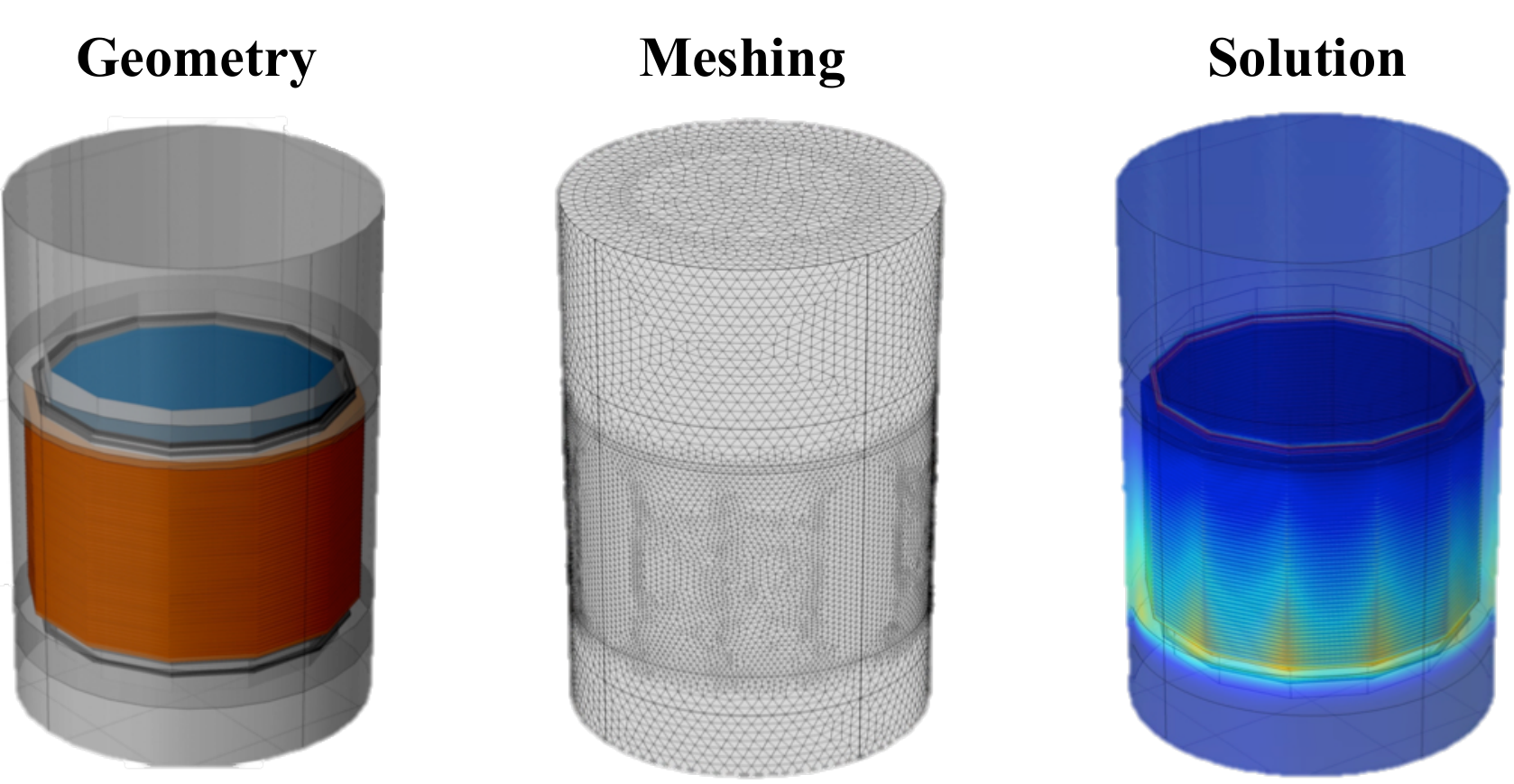}
\caption{Steps in building the 3D model of the electric fields in the LUX detector using \textsc{COMSOL} Multiphysics v5.0.}
\label{fig:comsol}
\end{center} 
\end{figure}

Once the detector geometry, materials, electrical dielectric properties, and voltages were assigned to detector volumes and boundaries, this defined the electrostatic conditions. \textsc{COMSOL} then generated a mesh by discretizing the space into tetrahedra and a proposed field map was generated. Using an adaptive mesh refinement allowed further improvement of mesh quality while minimizing solution error. This process is illustrated in figure~\ref{fig:comsol}. 

The resulting electric field values found in \textsc{COMSOL} Multiphysics were exported and used to produce a simulation dataset of electron-like particles in Python v2.7. This was accomplished by starting electron-like test particles uniformly throughout the detector's active volume and recording their intersection with the liquid surface after drifting. These particles were propagated step-wise in $1-3~\mu$s intervals using interpolated velocity since the electron drift velocity in LXe varies with electric field~\cite{Albert:2016bhh}. 

Once the electron-like test particles reached the liquid surface, the simulated drift time ($t_{sim}$) and location of the simulated S2 light production (\textit{x$_{\textrm{S2}_\textrm{sim}}$, y$_{\textrm{S2}_\textrm{sim}}$}) were compared with the (\textit{x\subStwo, y\subStwo, t}) distribution of \krm calibration data. See section~\ref{sub:charge_modeling} for details.

\krm was injected weekly into the detector as a calibration source producing $\sim10^{6}$ events that distribute homogeneously in the detector's active volume within minutes. This makes it a perfect source to use in developing 3D maps of the electric field. The radioactive isotope \krm has a half-life of 1.83~hours. The decay occurs in two transitions of 32.1 and 9.4 keV respectively, with an intervening half-life of 154 ns. These two transitions can each proceed according to multiple decay channels, with a high probability of internal conversion followed by Auger emission, resulting in a high concentration of decay energy into electron modes. More details regarding \krm calibration of the LUX detector can be found in~\cite{scott_krm}.

\subsection{Modeling charges in the LUX detector}\label{sub:charge_modeling}
\begin{figure}[ht!]
\begin{center}
\includegraphics[width=0.6\textwidth,clip]{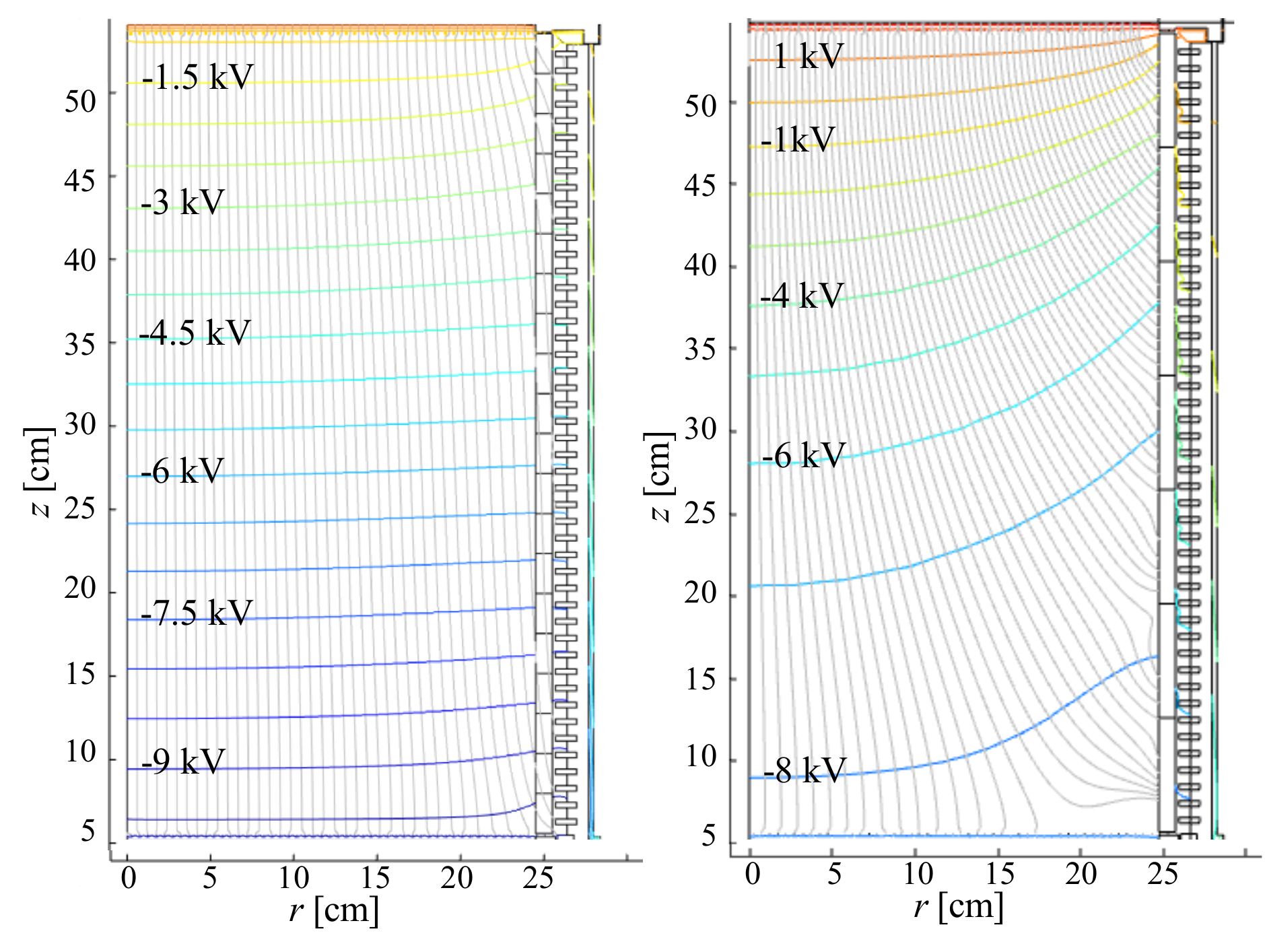}
\caption{Two simple axially symmetric \textsc{COMSOL} Multiphysics models built for visualization purposes. Left: electric field lines and equipotentials in the LUX detector during WS2013. A radially-inward component is seen, resulting from the geometry of the field cage and the grids. Right: electric field lines implied by \krm from 2015-04-26 showing the much stronger radially-inward component modeled by negative charge densities in the PTFE panels.}
\label{fig:field_lines}
\end{center} 
\end{figure}

A large change in the detector's electric field occurred between WS2013 and WS2014-16 which affected electron paths as shown in figure~\ref{fig:contour_runs34}. This change in field can be modeled with negative charge densities present in the PTFE panels; the consequence of this accumulated charge is illustrated in figure~\ref{fig:field_lines}.

\begin{figure}[ht!]
\begin{center}
\includegraphics[width=.85\textwidth,clip]{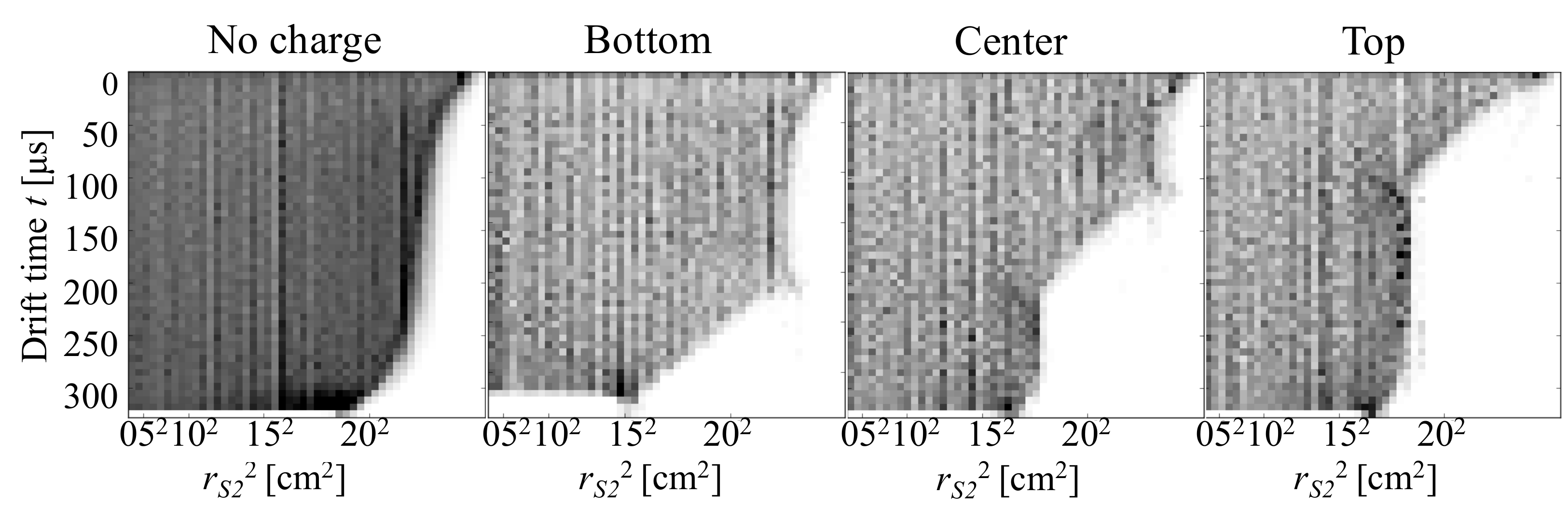}
\caption{Effect of negative charge densities in the PTFE panels on events in the detector. The 4 figures illustrate \textsc{COMSOL} models with no charge in the PTFE panels and with $-1~\mu$C/m$^2$ deposited at the lower, middle, and upper segments of a PTFE panel causing a strong inward push for events that originate near or below the charge site.}
\label{fig:charge_demo}
\end{center} 
\end{figure}

To gain intuition about the effects on electron paths caused by charge in the PTFE panels, a simple simulation was performed by depositing $-1~\mu$C/m$^2$ of charge density in either the lower, middle, or upper third of a PTFE panel; the resulting force causing inward displacement of events that originate near or below the charged site is demonstrated in figure~\ref{fig:charge_demo}.

To model the detector behavior in WS2014-16 first the basic charge-free model was modified to include WS2014-16 grid voltages. Then 42~additional \textsc{COMSOL} Multiphysics models were built with voltages on all boundaries set to 0~V. The 12 PTFE panels surrounding the active volume were split into 42~tiles by defining 6~angular and 7~vertical divisions of equal width and height. Therefore there were 2 PTFE panels in each angular division. Each one of the 42 models included $-1~\mu$C/m$^2$ of charge on one of the 42 tiles. These 43 datasets (arising from 42 models with charge and one without charge) served as basis vectors since the charge on any one tile could be scaled up or down as needed. Leveraging the superposition principle, these models linearly combined to produce electric field maps of arbitrary charge distribution. Those models were solved by \textsc{COMSOL} Multiphysics and the resulting 43 electric field vector maps were exported on a 1~cm$^3$ grid to a Python script where all further calculations were performed. For a given hypothetical charge distribution in the PTFE panels a simulation dataset was made and compared to \krm calibration data. This charge was varied with an iterative algorithm until the distributions converged on a stable solution.

Unphysical behavior emerged when 2 neighboring vertical tiles contained very different charges. To mitigate these sharp discontinuities the charge on each tile was set to fall off linearly both in the azimuthal and in \textit{z} directions. This smoothing was defined such that in the azimuthal direction each tile had a uniform $-1~\mu$C/m$^2$ charge density from 1/4 to 3/4 of tile width. This charge density then decreased linearly to $0~\mu$C/m$^2$ through 1/4 of the width of the neighboring tile. Therefore the charge density in the azimuthal direction was continuous. The same technique was applied in the \textit{z} direction except that the bottommost and topmost edge tiles were smoothed in only the direction pointing toward the center of the panel.

After the simulated particles were drifted, a likelihood function was used to compare \krm data to simulation. Then the initial charge hypothesis was refined and this approach was iterated using the Metropolis-Hastings algorithm until convergence of results was seen. The Metropolis-Hastings algorithm is a Markov Chain Monte Carlo method for sampling from a multi-dimensional distribution with a high number of degrees of freedom. First an initial guess was made for the amount of charge on each tile and the corresponding model was created. Each subsequent guess was randomly sampled from a Gaussian distribution centered at each charge density with a width of $\sim0.1~\mu$C/m$^2$. The ideal step size was tuned to achieve acceptance of new charge distributions of $\sim23\%$~\cite{chib1995understanding}, where the acceptance ratio was defined as
\begin{eqnarray}
\alpha=\frac{\mathcal{L}_{new}}{\mathcal{L}_{old}}.
\end{eqnarray}
Here, $\mathcal{L}_{new}$ and $\mathcal{L}_{old}$ are the log likelihood functions for the current and previous steps, respectively. To speed up the convergence at the beginning of the modeling efforts, the amount of charge in the PTFE panels was adjusted manually to smooth out charge deviations between neighboring vertical panels before using the Metropolis-Hastings algorithm. As the dataset neared convergence a modified Metropolis-Hastings algorithm was used due to the high number of degrees of freedom (42). Instead of varying all charges simultaneously, as is usual for the algorithm, only one random charge was varied at a time which sped up convergence.

\begin{figure}[!ht]
\begin{center}
\includegraphics[width=0.85\textwidth,clip]{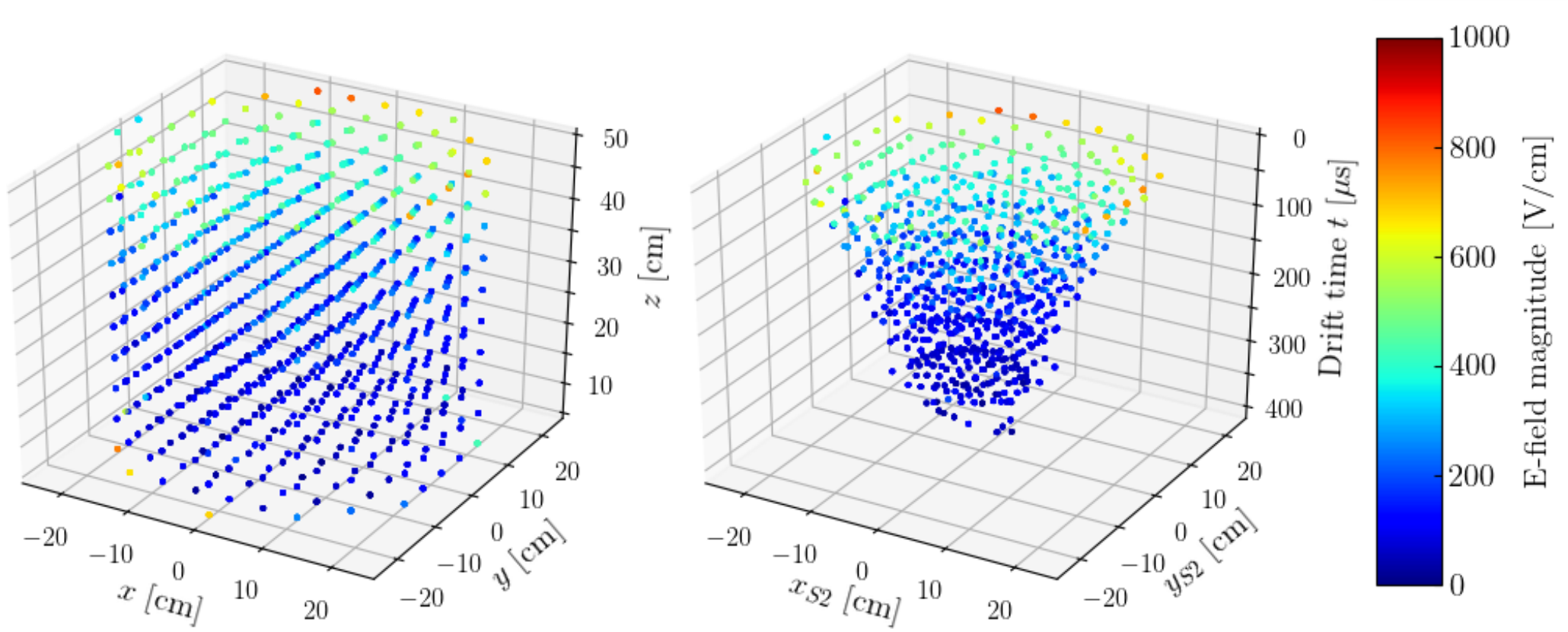}
\caption{Transformation from the reconstructed S2 space (right) to real space (left). The color scale shows the electric field strength at each point using \krm data from 2015-08-26.}
\label{fig:real_S2}
\end{center} 
\end{figure}

Once the final charge distribution was found, the information included in the resulting field maps was provided on a 1~cm$^3$ grid that included the detector real space coordinates (\textit{x, y, z}), the electric field in each location, and the coordinates as seen by the detector (\textit{x\subStwo, y\subStwo, t}). The transformation between real space and the reconstructed S2 space can be seen in the two plots in figure~\ref{fig:real_S2}. The field maps were used in a Geant4-based LUXSim~\cite{Akerib201263} simulation, position reconstruction, and limits code.

\subsection{Alternative explanations for the field distortion}
Other than the charge evolution in the PTFE surface, there are other plausible explanations for distorted fields inside a TPC. One possibility would be that drifting electrons land on the PTFE surface and do not get pulled toward the surface due to the high resistivity of PTFE. This is unlikely since the detector is designed to pull electrons up and away from the PTFE surface. Other explanations might be a short or open circuit in the field-shaping ring resistor chain. The field distortion caused by a shorted resistor was simulated in a simplified axially symmetric \textsc{COMSOL} model. A shorted resistor produces a small, very localized field deformity, unlike the detector-wide field distortion observed in WS2014-16. This effect would also not evolve throughout the detector operation. 

An open circuit was also simulated in the axially symmetric \textsc{COMSOL} model. A field distortion is obvious and not unsimilar to the field observed in WS2014-16, however depending on the location of the open field, it can create large, dead, field-less regions in the detector. The resistance of the field-shaping ring resistor chain was measured before WS2013 and then again before WS2014-16. No significant change was observed, ruling out this explanation. An open circuit effect is also ruled out because it would not create a time-dependent component. Additionally, \isot{Cs}{137} calibrations were performed in WS2014-16 using acrylic radioactive-source deployment tubes suspended along the side of the detector~\cite{Akerib:2012ys}. These calibrations showed events along the entire detector height, inconsistent with the presence of field-less regions.

\section{Modeling detector observables}

To model the electric field, the simulation results were compared to the field implied by \krm calibration data to ensure its accuracy. This comparison had to be done many times during the Metropolis Hastings algorithm rendering the traditional method of drifting electrons through the detector computationally too intensive and an alternative method comparing event densities was developed. 

\begin{figure}[ht!]
\begin{center}
\includegraphics[width=0.7\textwidth,clip]{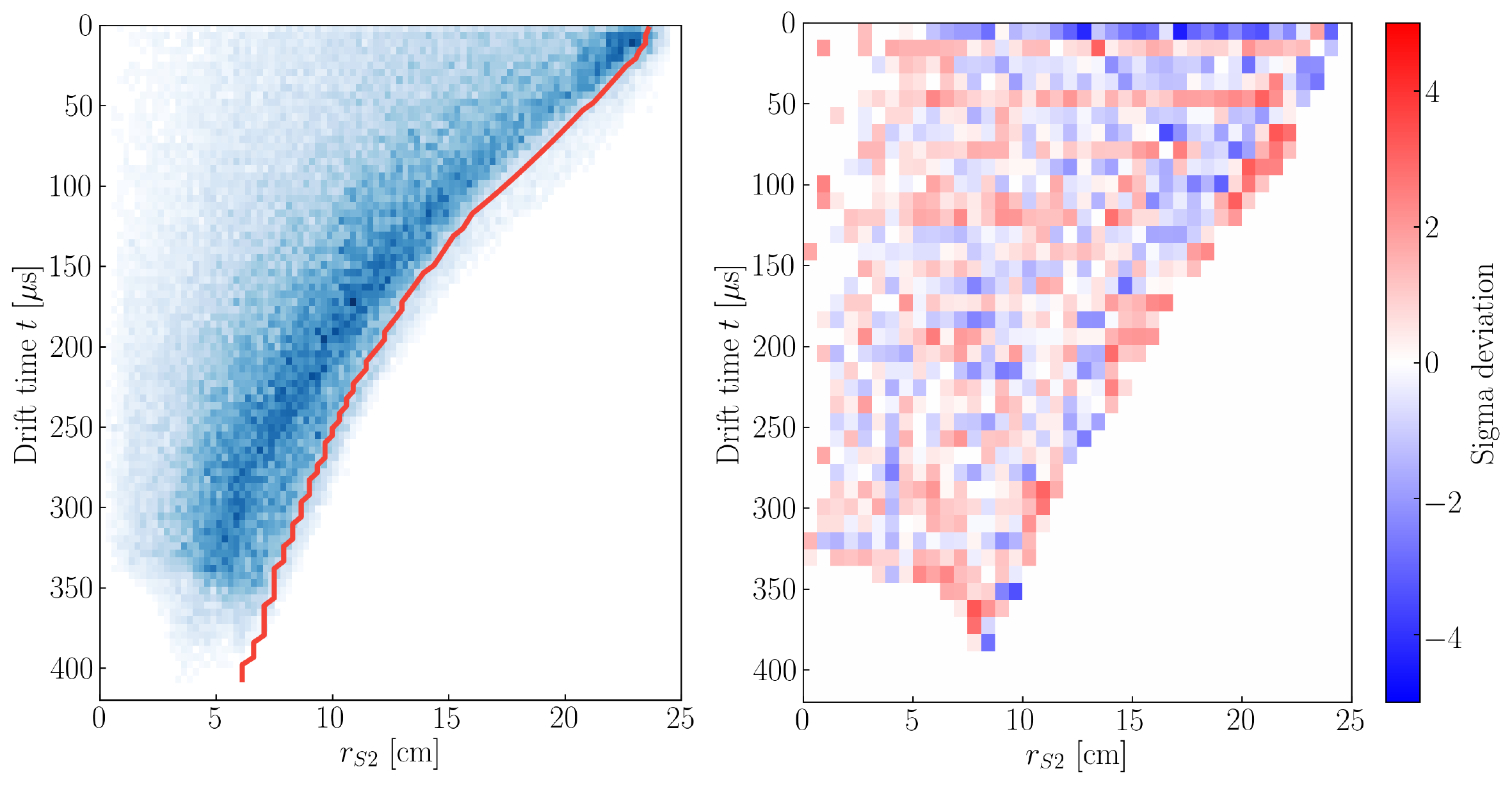}
\caption{Left: solid red contour at $r=22$~cm illustrates the modeled edge of the \krm distribution (blue) from 2016-02-22. Right: variation between normalized histograms of \krm data and simulation from 2016-10-06 weighted by the error in each bin (using an azimuthal slice between $120^{\circ} \leq \theta < 180^{\circ}$).}
\label{fig:sigma_fit}
\end{center} 
\end{figure}

To begin, the \krm events as seen in S2 space were reverse propagated through the modeled drift field to find the initial homogeneous distribution. In practice, this was accomplished by segmenting the \krm data into tetrahedra within the active volume using Delaunay triangulation and recording the number of events enclosed by each segment. The vertices of those tetrahedra were drifted ``back in time'' through the electric field; each vertex started at the liquid surface and was drifted through the active volume, making the transformation from S2 to real space. The new volume of each tetrahedron after drifting, $V_\textrm{tet}$, was calculated and the number of \krm events, \textit{n\subStwo}, contained in each tetrahedra recorded. Since in real space \krm events were uniformly distributed throughout the detector, the number of events in each tetrahedron was proportional to its volume $n_\textrm{real} = N_\textrm{Kr}*\frac{V_\textrm{tet}}{V_\textrm{TPC}}$, where $N_\textrm{Kr}$ is the total number of \krm events in the detector and $V_\textrm{TPC}$ is the active volume of the detector. If the simulated electric field captured the field in the detector well, then $n_\textrm{real}$ and \textit{n\subStwo} had equal number of events. The density comparison between \krm data and simulation is illustrated in the right panel of figure~\ref{fig:sigma_fit}.

To complete the fine-tuning of the model, the edges of the \krm distribution were fully simulated to compare with data, in a method independent of the \krm homogeneity in the detector. When approaching the final convergence of data and simulation only the edges of the \krm distribution were used to fine tune the model. The detector was split into 12 angular sections corresponding to the 12 PTFE panels and negative charges were drifted to the liquid surface starting along the center of each panel face. An offset radius (set at $r=22$~cm) was used due to a higher fidelity of field maps slightly inward from the PTFE panels. The resulting simulated edge at the detector wall ($r=23.7$~cm) agrees with the edge found both from the \krm data and from the location of the wall as determined from {\isot{Po}{210}} background~\cite{bg_paper}. It is worth noting that even though only the detector wall was fitted, as shown in the left panel of figure~\ref{fig:sigma_fit}, the result correctly described the density distribution inside the entire detector.

\subsection{WS2013 modeling results}

\begin{figure}[ht!]
\begin{center}
\includegraphics[width=0.7\textwidth,clip]{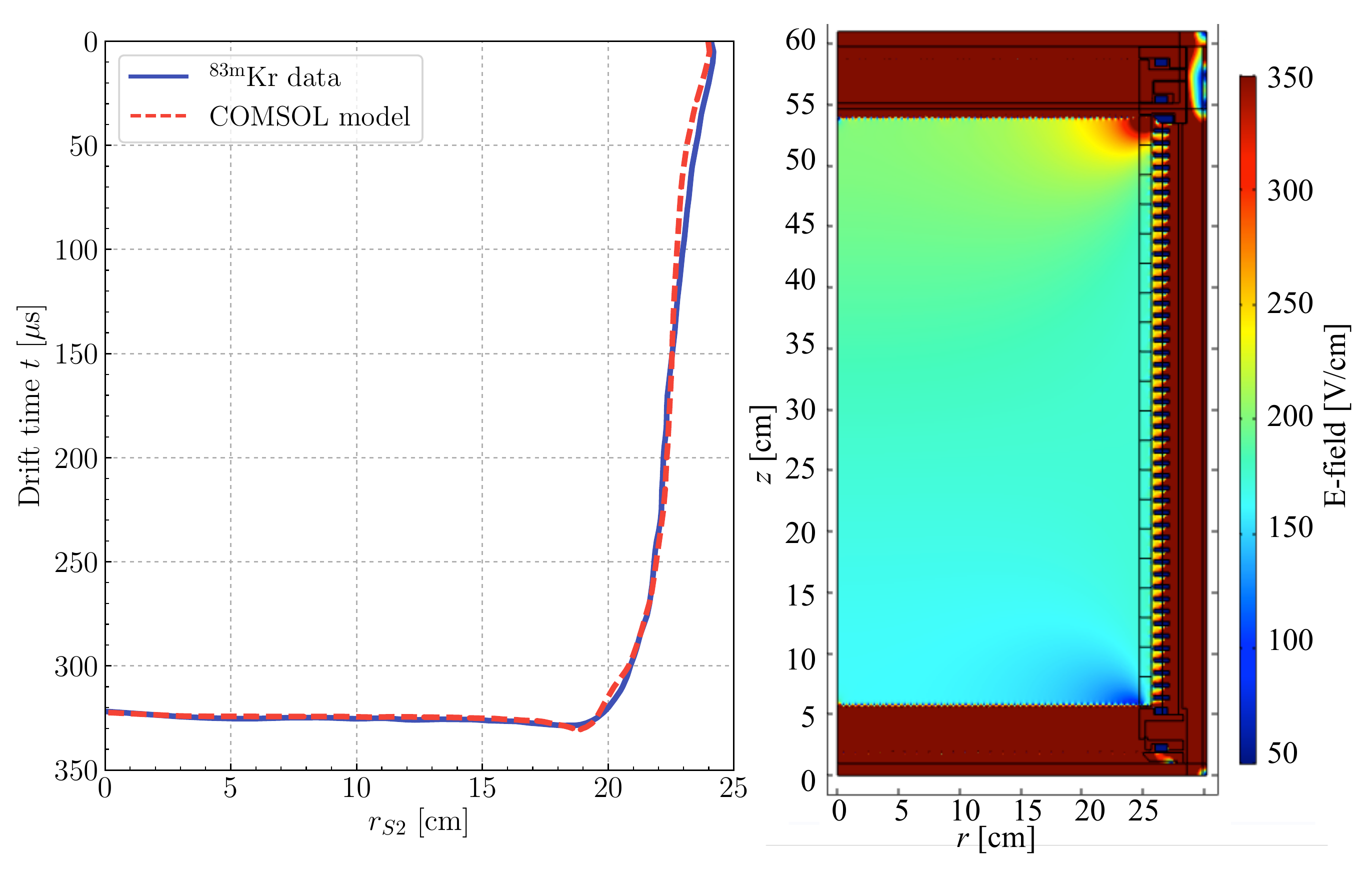}
\caption{Left: contour plot showing histogram edges of the 3D field model and \krm events averaged over all azimuthal angles during WS2013. Data (solid blue) and simulation (dashed red) contours were created following the method described in figure~\ref{fig:contour_runs34} without any fitting or tuning necessary. Right:~analogous electric field strength in LUX during WS2013. Regions in the corners of the detector show field leakage through the transparent boundary grids.}
\label{fig:run3}
\end{center} 
\end{figure}

A comparison of the 3D field model and \krm data from WS2013 is shown in the left panel of figure~\ref{fig:run3}. Excellent agreement is seen, without a need to tune any aspect of the model to improve agreement with the data. The value of the mostly uniform electric field inside the fiducial volume during WS2013 was $177\pm14$~V/cm as simulated in \textsc{COMSOL} Multiphysics. In WS2013 the fiducial volume was defined by drift time range of $38-305~\mu$s and radius $r\leq20$~cm, and contained 118~kg of LXe~\cite{Akerib:2015rjg}. The electric field strength in WS2013 is illustrated in the right panel of figure~\ref{fig:run3}.

The slight curve at high drift time in the reconstructed (S2) coordinates is due to the field transparency of the cathode, causing a leakage field from below the cathode~\cite{mei2012direct}. In WS2013 this effect was corrected in the analysis without a need for a field map~\cite{scott_krm}. A similar effect caused by the transparency of the gate grid can be seen at low drift times near the gate grid. These field leakages introduce a radial field component to the field lines as shown in figure~\ref{fig:field_lines}, illustrating that an electron originating at high radius just above the cathode follows the field lines shown and reaches the liquid level at a radius reduced by several centimeters compared to the initial position. 

\begin{figure}[ht!]
\begin{center}
\includegraphics[width=0.6\textwidth,clip]{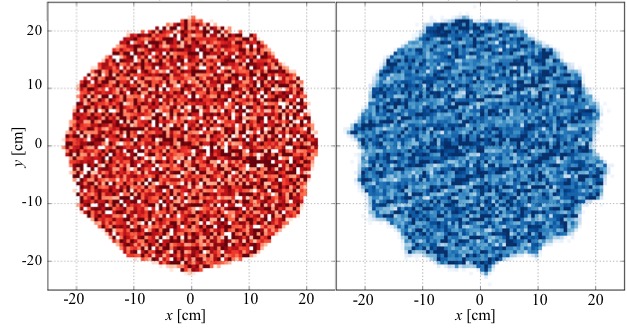}
\caption{Left: original \textsc{COMSOL} model with no charge showing a clear dodecagonal cross-section. Right: data from the detector with visible scallops using \krm data from 2013-05-10 during WS2013. Also visible is the orientation of gate wires. Both cross-sections are a slice at a drift time of $255\,\mu$s $ < t \leq 290\,\mu $s.}
\label{fig:scalloping}
\end{center} 
\end{figure}

Using the tiled charge model developed to study the electric field for WS2014-16, we were able to qualitatively improve our understanding of the WS2013 field model with respect to the detector wall shape. Even though the electric field in WS2013 was mostly uniform, a scalloping was seen along the edges of the detector that was not observed in the ideal \textsc{COMSOL} Multiphysics model without charge tiles, as shown in figure~\ref{fig:scalloping}.

The same approach described in section~\ref{sub:charge_modeling} was used to model this scalloping effect in WS2013. Since the tiling was optimized for WS2014-16, the model was unable to fully reproduce all the fine features. The average fitted charge found in the PTFE panels was $-0.03~\mu$C/m$^2$, which is comparable to triboelectric charge densities. This illustrates the sensitivity of the field modeling method to capture even very small charge densities.

\subsection{Alternative determination of electric field magnitude using NEST}
An alternative determination of the electric fields was done by fitting the field as a free parameter in the Noble Element Simulation Technique (NEST) package~\cite{Szydagis:2013sih} to match calibration data. NEST encompasses the WIMP search community's understanding of both scintillation and ionization yields as well as the field impact on the electron-ion recombination probability. As the field magnitude is increased, the ionization signal increases while the scintillation signal decreases~\cite{Aprile:2006kx}. This is a result of the electric field influencing the fraction of electrons which recombine with ions at a particle-interaction site: if a given electron recombines with an ion, it contributes to the scintillation signal, otherwise it contributes to the ionization signal. NEST's underlying physics model accounts for these electron-ion recombination effects, and has been refined with recent measurements in the literature~\cite{Lin:2015jta} to extend up to high fields.

Because NEST requires an electric field magnitude as input, the field value can be treated as a free parameter and NEST's output fit to calibration data. By performing such a fit to calibration data from a specific region of the detector, an estimate of the field magnitude in that region at the time of the calibration was obtained. This technique for estimating field is completely independent from the technique described in previous sections: this technique uses the size of the scintillation and ionization pulses, while the technique described previously in section~\ref{Modeling_Efields} uses only the spatial distribution of the pulses.

Though \krm decays were used for the field estimates in previous sections, this source is not conducive to analysis with NEST for this purpose because of its complicated LXe response~\cite{Baudis:2013cca} due to its decay~\cite{Chan:1995tk}, which is difficult to model. Instead, periodic calibrations with tritiated methane (CH$_3$T) dissolved in the LXe were used. The decay of this source is very simple and covers a wide range of energies~\cite{Akerib:2015wdi}. Data collected from this source form a band in the space of $\log_{10}(\mathrm{S2/S1})$ vs.~S1, whose mean and width depends on the applied field. The spatial variation in detector response was handled by dividing the fiducial volume into four regions of drift time; the boundaries of these regions were 40, 105, 170, 235, 300~$\mu$s.

CH$_3$T data were segregated into these four regions, and the band from each was considered independent. Four spatial regions were found to be the optimal number: more regions would result in too few calibration statistics with which to tune the model for each region, while fewer regions would not adequately capture the field variation in the full fiducial volume.

\begin{figure}[hb!]
\begin{center}
\includegraphics[width=0.8\textwidth,clip]{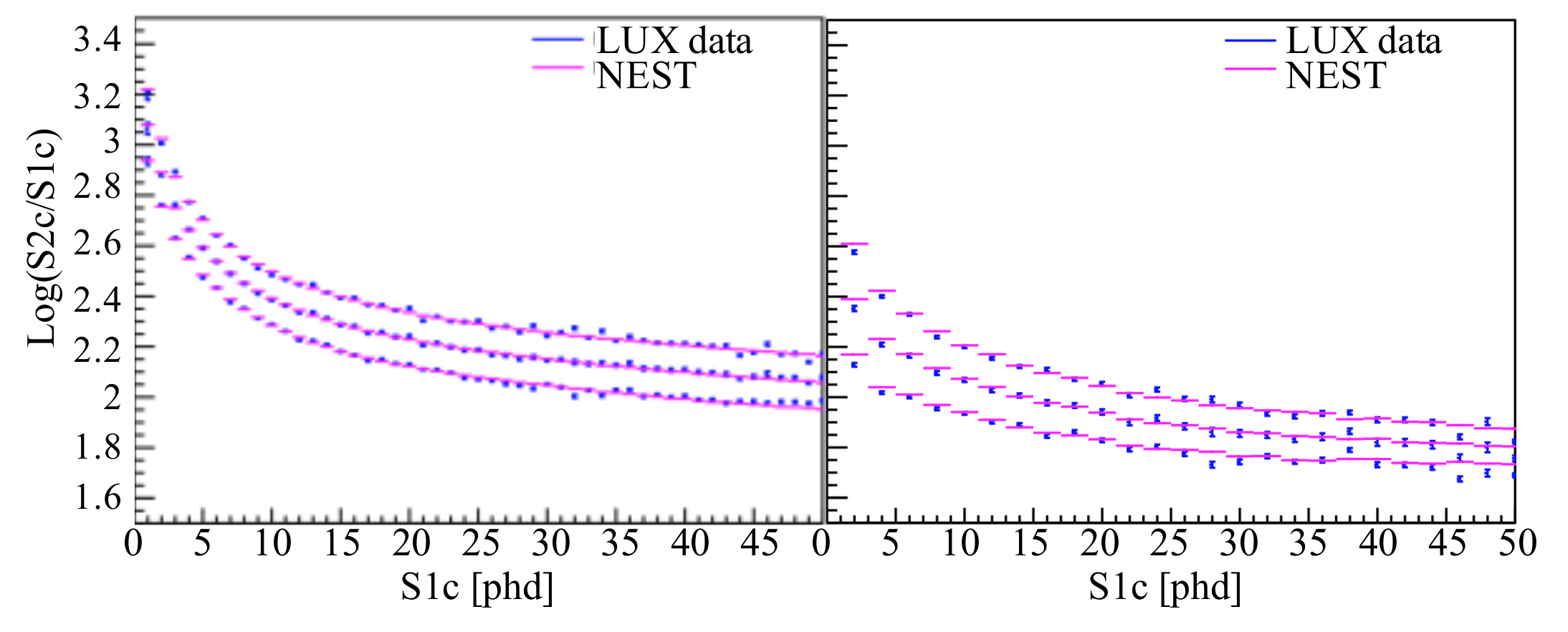}
\caption{Agreement between LUX data and NEST simulation for the ER band (left) and NR band (right) along with the 10\% and 90\% confidence level bands. Data shown are from September 11, 2014 - January 1, 2015 period with drift time $40~\mu$s~$ <~t~<~105~\mu$s as defined in~\cite{Akerib:2016vxi}.}
\label{fig:NEST}
\end{center} 
\end{figure}

The CH$_3$T band from each drift-time region was broken up into bins of S1; for each S1 bin, the distribution of events in $\log_{10}(\mathrm{S2/S1})$ was fit with a normal distribution. The CH$_3$T band as a function of S1 was characterized by the means and widths of these fits, and these were compared to the means and widths predicted by the output of NEST. Two NEST parameters were varied until optimal agreement between NEST and CH$_3$T data was obtained; these parameters were (1) the applied field (which affects mainly the mean value of the electron-ion recombination fraction), and (2) a parameter, $f_r$ that characterizes the size of the fluctuations in this electron-ion recombination. Because the electric field is considered uniform within a given fiducial-volume region, residual variations in the field were empirically accounted for by increasing $f_r$. Figure~\ref{fig:NEST} (left) shows the comparison between data and best-fit model for CH$_3$T in one of the four spatial regions at the beginning of the WIMP-search run. Data and best-fit models for all four drift-time regions, at all times during the WIMP-search run are consistent. The quantities in the axis labels are given as ``S1c'' and ``S2c'', which indicate that the pulse areas have been corrected for variations resulting from non-field-related effects such as xenon purity and geometrical optics.

Because LXe's response to nuclear recoils is significantly different than for electronic recoils, nuclear recoils require a separate detector-response model than is obtained from the CH$_3$T calibrations. For this purpose, data collected with a collimated beam from a deuterium-deuterium (DD) neutron generator were used~\cite{Akerib:2016mzi}. However, LXe's nuclear-recoil response is nearly unaffected by applied field~\cite{Aprile:2006kx}, and therefore the NEST models that are applied to the DD calibrations use the field magnitudes as determined by the corresponding CH$_3$T calibration of the same spatial region and date. The fits of the NEST model to DD data therefore vary only the parameter $f_r$. Figure~\ref{fig:NEST} (right) shows the comparison between bands from DD data and best-fit model for the same spatial region and date as the CH$_3$T band seen in the left plot.

\section{Time-dependence of electric fields during WS2014-16}

\subsection{Charge evolution}\label{sub:charge_evolution}

\begin{figure}[ht!]
\begin{center}
\includegraphics[width=0.6\textwidth,clip]{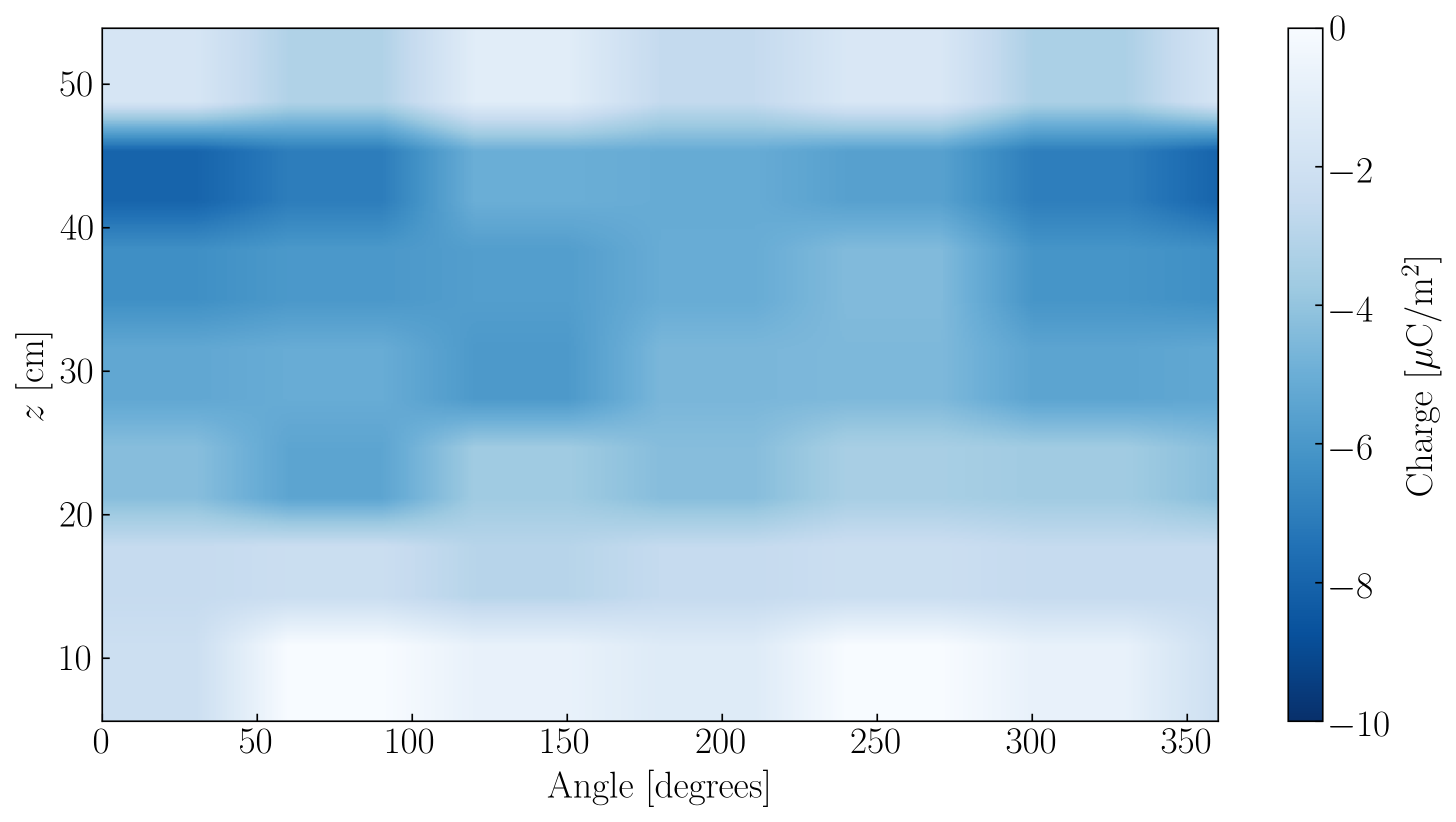}
\caption{Charge density distribution in the detector's PTFE panels that resulted from fitting \krm data from 2014-10-06. The 42 segments correspond to the smoothed tile charges as described in section~\ref{sub:charge_modeling}.}
\label{fig:charge_2d_1410}
\end{center} 
\end{figure}

\begin{figure}[h!]
\begin{center}
\includegraphics[width=0.6\textwidth,clip]{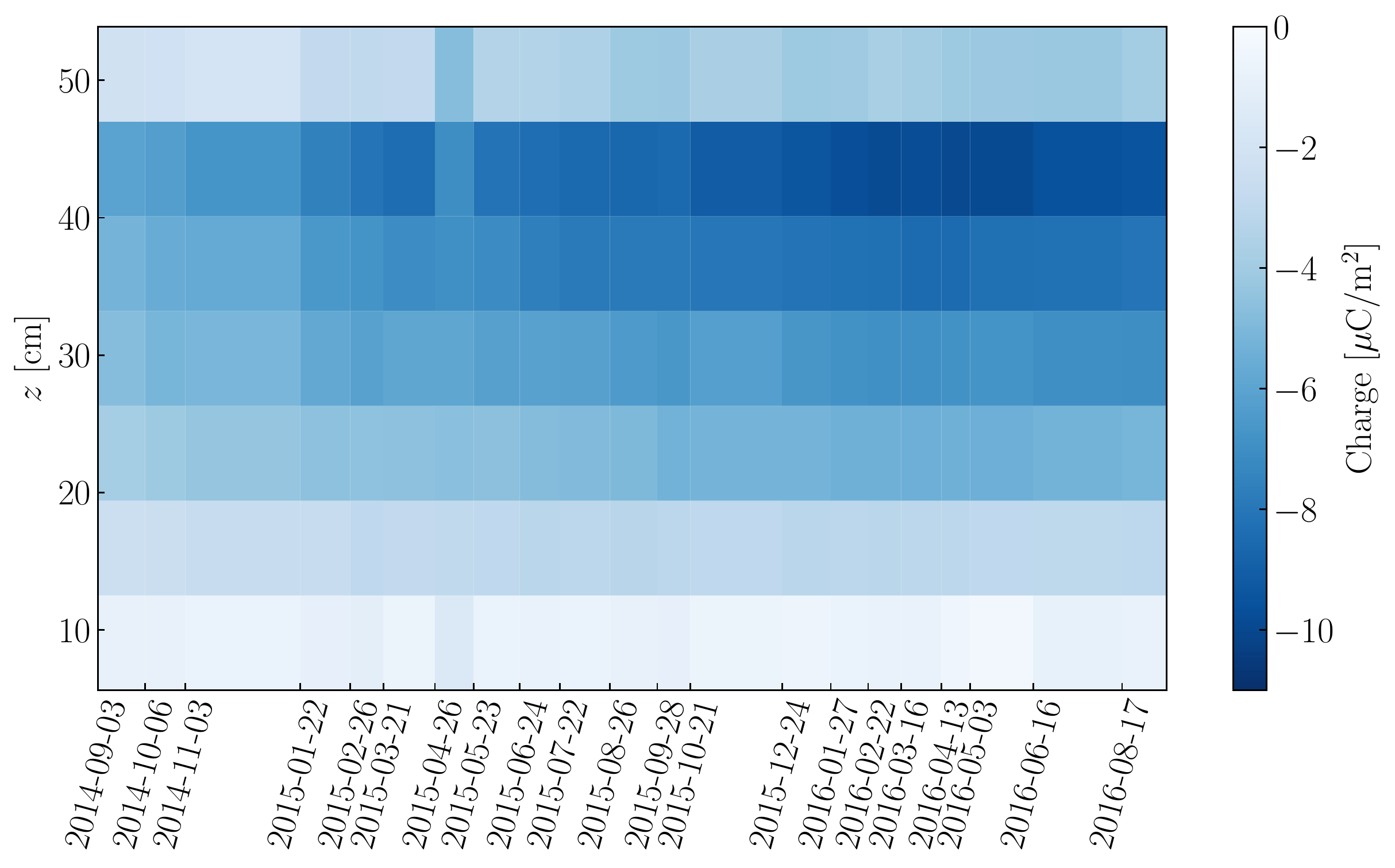}
\caption{Charge density in each \textit{z} segment averaged over all azimuth throughout WS2014-16.}
\label{fig:charge_time_z}
\end{center} 
\end{figure}

\begin{figure}[h!]
\begin{center}
\includegraphics[width=1.\textwidth,clip]{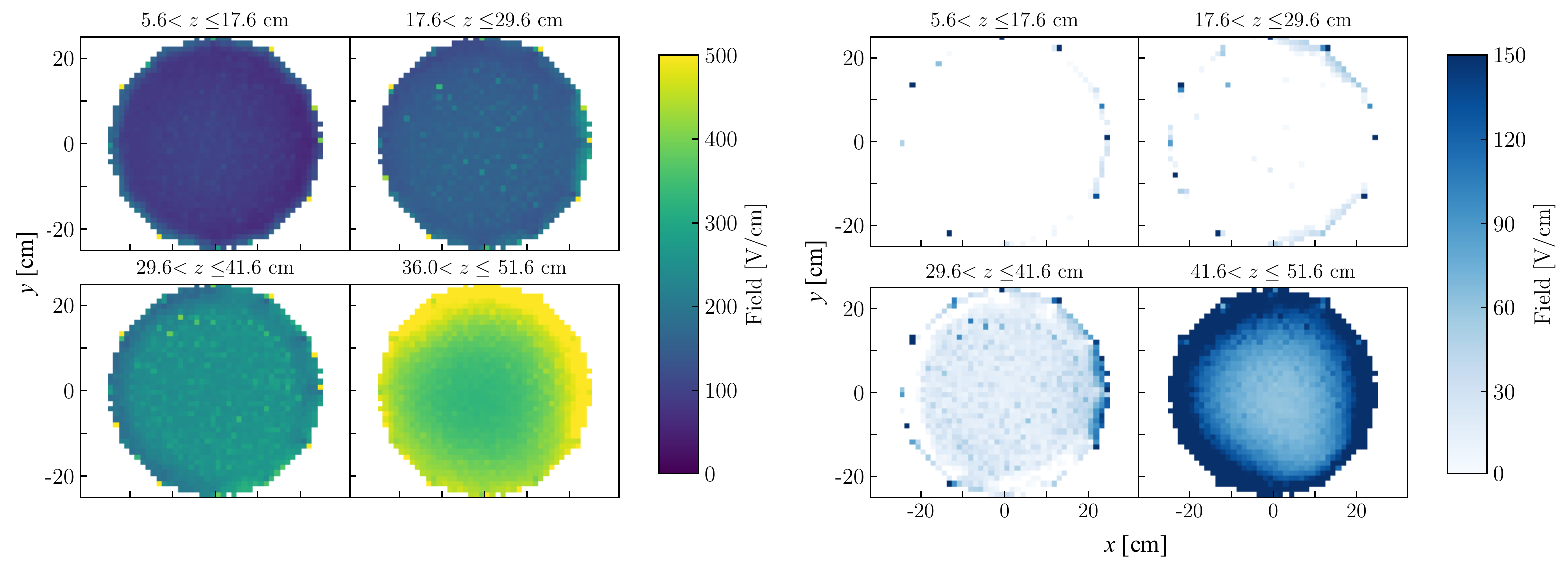}
\caption{Left: electric field in different \textit{z} segments of the detector as modeled for 2014-09-03 dataset illustrating the azimuthal non-uniformity. Right: difference between the electric field at the end of WS2014-16 (field map from 2016-05-03) and the beginning (field map from 2014-09-03). The change in field is not uniform throughout WS2014-16.}
\label{fig:field_diff}
\end{center} 
\end{figure}

The results of the method described here were one field map for each month of WS2014-16, corresponding to 21 field maps in total. As discussed earlier, each field map was produced by adjusting the negative charge distribution in the PTFE panels and fitting the modeled electric field to the detector observables. One of the resulting charge distributions in the PTFE panels is shown in figure~\ref{fig:charge_2d_1410} and the evolution of charge over the course of WS2014-16 is demonstrated in figure~\ref{fig:charge_time_z}. The varying charge results in a time dependent electric field, changing also in azimuth and \textit{z} as illustrated in figure~\ref{fig:field_diff}. Most negative charge was found to be in the top 1/3 of the detector throughout all of WS2014-16, possibly due to electrons in the PTFE panels being pulled toward the gate while holes were attracted to the cathode at a faster rate. 

\begin{figure}[ht!]
\begin{center}
\includegraphics[width=0.55\textwidth,clip]{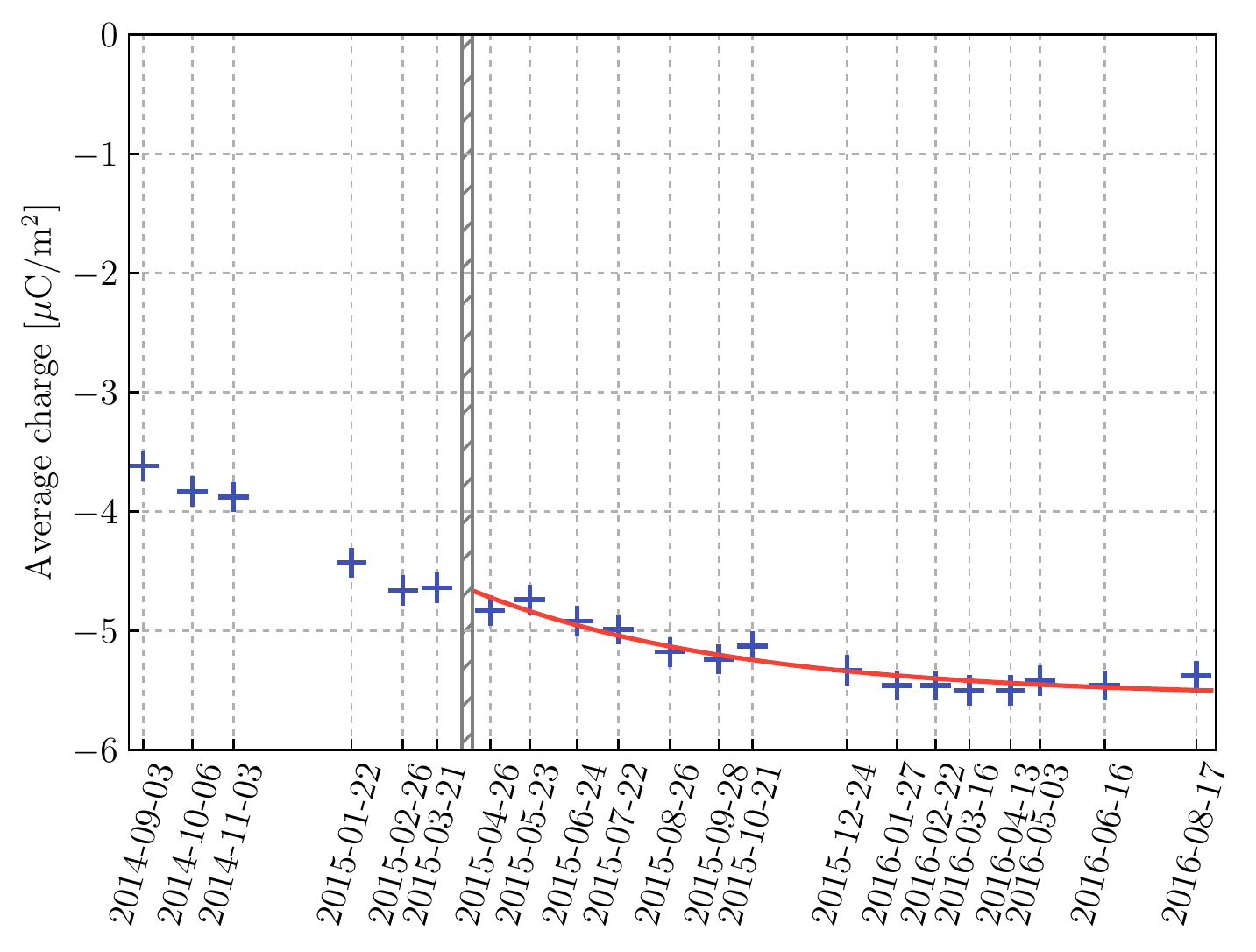}
\caption{Increasingly negative charge density obtained from the field models, averaged over the entire PTFE panel surfaces for each month during WS2014-16. The data points correspond to the modeled average charge densities in the PTFE panels obtained from a small subset of the regular \krm datasets. These data points can be fitted with an exponential function $\sigma=3.1\exp(-t/181)-5.6$ where \textit{t} is in units of ``days since 2014-09-03''. The gray dashed line indicates a week in April 2015 when the cathode voltage was biased down to 0~V to observe its effect on the charge.}
\label{fig:avg_charge_fit}
\end{center} 
\end{figure}

The increasingly negative charge throughout WS2014-16 averaged over the entire PTFE area is plotted in figure~\ref{fig:avg_charge_fit}. The cathode voltage was biased down to 0~V from April~7 to April~15, 2015 to investigate its effect on the charge in the PTFE panels. Data from this period was not used in analysis. After the cathode was re-biased to $-8.5$~kV the charge density was fit with an exponential function as in Reference ~\cite{mellinger2004charge}:
\begin{eqnarray}
\sigma=\sigma_{0}\exp(-t/\tau)+\sigma_\infty\label{eq:charge_fit}
\end{eqnarray}
where $\sigma$ is the average surface charge density, $\sigma_{0}$ is the initial charge density, and $\sigma_\infty$ is the value of the charge $\lim_{t\to\infty}\sigma$. $\tau$ is the charge transit time. The best-fit values found were $\sigma_{0}=3.1\mu$C/m$^2$, $\tau=181$~days, and $\sigma_\infty=-5.6\mu$C/m$^2$.

There is a lack of research of PTFE properties at LXe temperatures so it is difficult to compare this number to the literature due to different experimental conditions used, such as temperature or the electric field applied. 

\subsection{Electric fields in WS2014-16 analysis}
\begin{figure}[h!]
\begin{center}
\includegraphics[width=1\textwidth,clip]{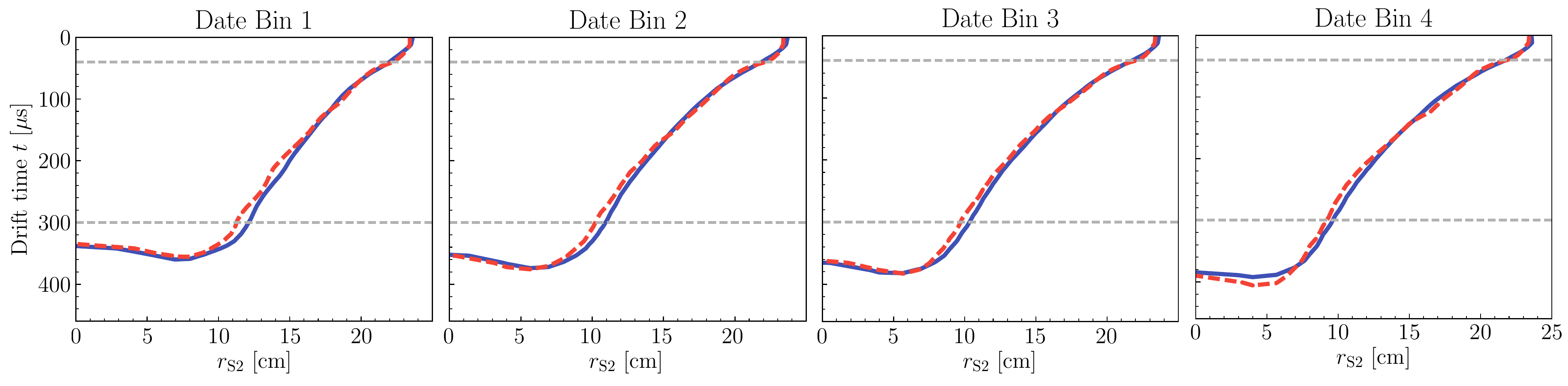}
\caption{A comparison of the measured position of the detector wall and the cathode from \krm (solid blue) to that predicted by the best-fit electric field model (dashed red). The contours were created following the method described in figure~\ref{fig:contour_runs34}. Horizontal gray dashed lines, at 40 and 300~$\mu$s, indicate the drift time extent of the fiducial volume used in WS2014-16.}
\label{fig:contour_time_bins}
\end{center} 
\end{figure}

For simplicity of the analysis in~\cite{Akerib:2016vxi} the data were divided into 16 bins, within each of which the change in the wall position was not significant. There were four bins in time, bounded by the following dates: September 11, 2014; January~1, 2015; April~1, 2015; October~1, 2015; May~2, 2016. Each of those bins was further split into four drift time bins with boundaries of 40, 105, 170, 235, and 300~$\mu$s. Only one field map was chosen for each date bin, hence only four field maps were used in WS2014-16 analysis. Each map was chosen to be close to the average of the electric field values in the segment. The agreement between each field map and the given \krm distribution for each date bin is shown in figure~\ref{fig:contour_time_bins}.

\begin{figure}[ht!]
\begin{center}
\includegraphics[width=0.7\textwidth,clip]{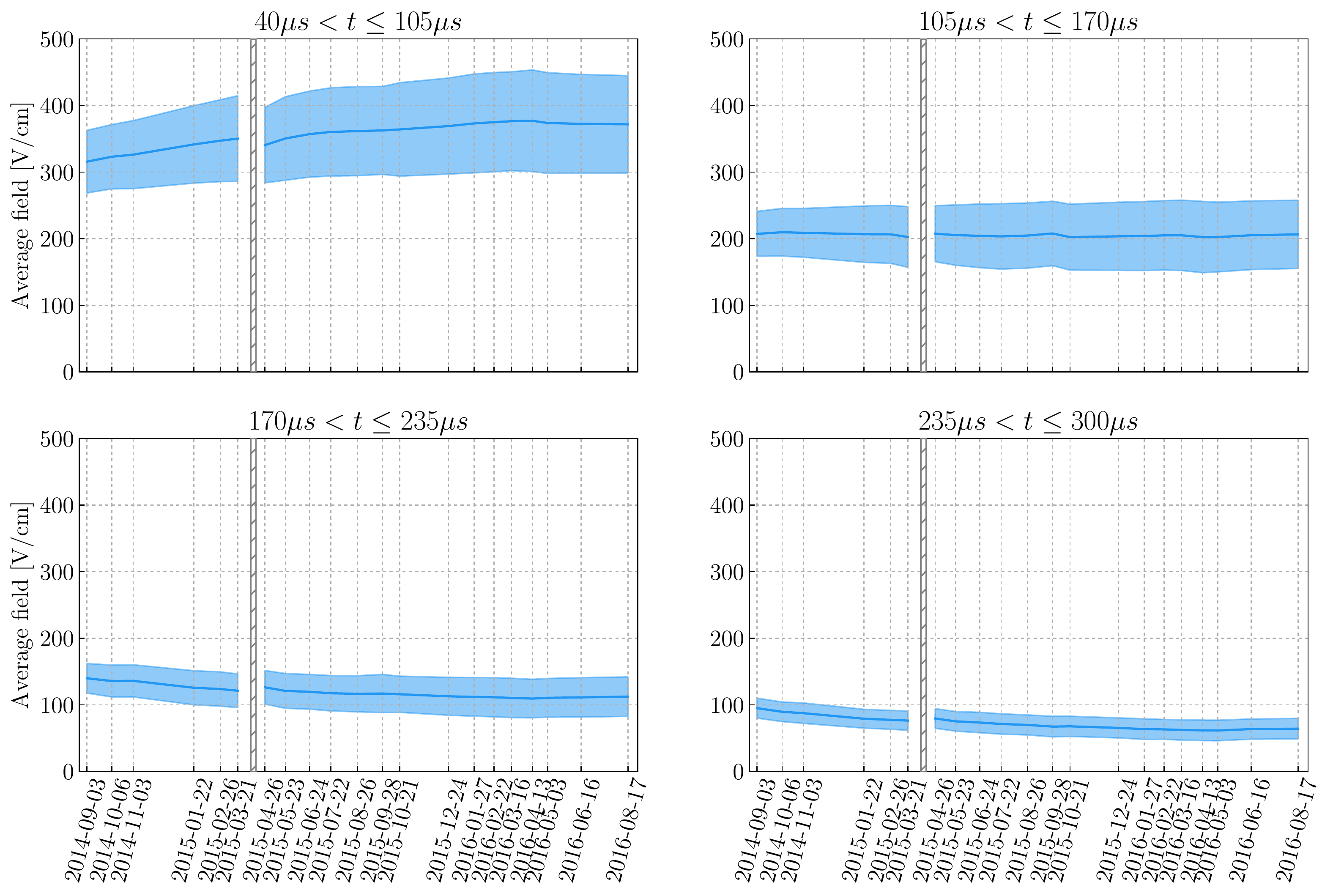}
\caption{Average field during WS2014-16 in the four drift time bins used in analysis. Bands indicate the standard deviation of the field in the given drift time bin. A radial cut at $r=20$~cm and a drift time cut at $40\,\mu$s $ < dt \leq 300\,\mu$s were applied. Gray dashed lines indicate a week in April 2015 when the cathode voltage was biased down to 0~V.}
\label{fig:avg_field}
\end{center} 
\end{figure}

Figure~\ref{fig:avg_field} shows the value of the electric field throughout the WS2014-16 analysis~\cite{Akerib:2016vxi}. It shows the divergence of the field extrema as the electric field magnitude varied from $\sim50-20$~V/cm near the bottom of the detector to $\sim500-650$~V/cm near the top. The mean value of the field of $\sim200$~V/cm remained mostly constant throughout the exposure.

\begin{figure}[ht!]
\begin{center}
\includegraphics[width=0.5\textwidth,clip]{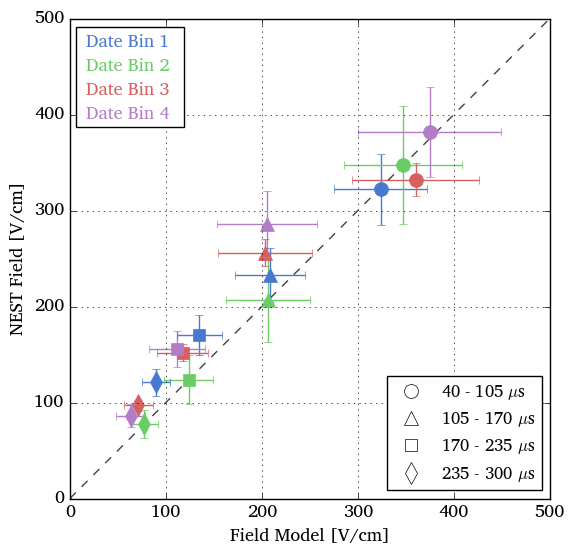}
\caption{Comparison of electric field as modeled using the COMSOL modeling method described in section~\ref{Modeling_Efields} to the best fit predicted by NEST for the 16 detector models (four spatial regions and four time periods) used in WS2014-16 analysis.}
\label{fig:nest_vs_field}
\end{center} 
\end{figure}

The volume-averaged fields obtained in \textsc{COMSOL} models can be compared to the NEST model fits that assume a uniform field within a given drift time bin. Figure~\ref{fig:nest_vs_field} shows the comparison of the field magnitudes obtained by the NEST fits (``NEST Field'') to the corresponding mean field values obtained from the \textsc{COMSOL} studies (``Field Model'') for all 16 detector models (four spatial regions and four time periods). The field estimates obtained from the two methods show good agreement. This point deserves emphasis, because these two techniques for estimating electric field magnitude are completely independent: the electrostatic field model is based on the observed electron drift paths alone, while the NEST fits are based on the S1 and S2 amplitudes alone. Therefore we have confidence in the S1 and S2 signal reconstruction techniques used in the LUX WIMP search analysis.

\section{Summary}
A method was developed for detailed modeling of the varying electric fields inside the active volume of the LUX detector throughout its scientific operations. Understanding of the electric field was vital for the ER and NR signal reconstruction, the foundation of the LUX WIMP search analysis. The fields observed in the bulk of the active volume of the detector during WS2013 were reproduced without any tuning necessary. The \textsc{COMSOL} field map modeling method described in this work is sensitive down to $-0.03~\mu$C/m$^2$, enough charge density to model a minor scalloping to the detector edge as seen in WS2013.

The time dependence of the electric field during WS2014-16 is likely due to the creation of charge in the PTFE panels during detector grid conditioning prior to the extended run. The model of electric fields was created by simulating the LUX detector in 3D in \textsc{COMSOL} Multiphysics, where hypothetical charges were deposited in the PTFE panels surrounding the active region and varied until convergence on the distribution of \krm calibration data within the detector was obtained. The resulting method allowed a faithful reconstruction of electric fields in the LUX detector and was critical for accurate event reconstruction and background discrimination during the WS2014-16 analysis~\cite{Akerib:2016vxi}. The fields in WS2014-16 varied in time, azimuth and depth and the method described in this work was able to capture this evolution in all dimensions. Twenty-one field maps were developed to accurately model the changing electric field throughout the extended WS2014-16 with an average modeled charge distribution in the PTFE panels varying from -3.6 to $-5.5~\mu$C/m$^2$.

The mean electric field values resulting from this \textsc{COMSOL} modeling method were found to be in good agreement with electric field values deduced from the NEST package, which fit the field-dependent scintillation and ionization yields of the CH$_3$T calibration source. The successful modeling of electric fields, along with the strong calibration and simulation program developed by the LUX collaboration enabled a thorough understanding of the LUX detector throughout all of the scientific program and strengthened its sensitivity to WIMPs.

\begin{acknowledgments}
This work was partially supported by the U.S. Department of Energy (DOE) under Award No.~DE-AC02-05CH11231, No.~DE-AC05-06OR23100, No.~DE-AC52-07NA27344, No.~DE-FG01-91ER40618, No.~DE-FG02-08ER41549, No.~DE-FG02-11ER41738, No.~DE-FG02-91ER40674, No.~DE-FG02-91ER40688, No.~DE-FG02-95ER40917, No.~DE-NA0000979, No.~DE-SC0006605, No.~DE-SC0010010, and No.~DE-SC0015535; the U.S. National Science Foundation under Grants No.~PHY-0750671, No.~PHY-0801536, No.~PHY-1003660, No.~PHY-1004661, No.~PHY-1102470, No.~PHY-1312561, No.~PHY-1347449, No.~PHY-1505868, and No.~PHY-1636738; the Research Corporation Grant No.~RA0350; the Center for Ultra-low Background Experiments in the Dakotas (CUBED); and the South Dakota School of Mines and Technology (SDSMT). LIP-Coimbra acknowledges funding from Funda\c{c}\~{a}o para a Ci\^{e}ncia e a Tecnologia (FCT) through the Project-Grant No.~PTDC/FIS-NUC/1525/2014. Imperial College and Brown University thank the UK Royal Society for travel funds under the International Exchange Scheme (No.~IE120804). The UK groups acknowledge institutional support from Imperial College London, University College London and Edinburgh University, and from the Science \& Technology Facilities Council for PhD studentships No.~ST/K502042/1 (AB), No.~ST/K502406/1 (SS) and No.~ST/M503538/1 (KY). The University of Edinburgh is a charitable body, registered in Scotland, with Registration No.~SC005336.

This research was conducted using computational resources and services at the Center for Computation and Visualization, Brown University, and also the Yale Science Research Software Core. The $^{83}$Rb used in this research to produce $^{83\mathrm{m}}$Kr was supplied by the United States Department of Energy Office of Science by the Isotope Program in the Office of Nuclear Physics.

We would like to thank the referees for their constructive comments and recommendations. We gratefully acknowledge the logistical and technical support and the access to laboratory infrastructure provided to us by SURF and its personnel at Lead, South Dakota. SURF was developed by the South Dakota Science and Technology Authority, with an important philanthropic donation from T. Denny Sanford, and is operated by Lawrence Berkeley National Laboratory for the Department of Energy, Office of High Energy Physics.

\end{acknowledgments}

\bibliographystyle{JHEP}
\bibliography{run4_fields}

\providecommand{\href}[2]{#2}\begingroup\raggedright\begin{thebibliography}{10}

\bibitem{Akerib:2013tjd}
{\scshape LUX} collaboration, D.~S. Akerib et~al., \emph{{First results from
  the LUX dark matter experiment at the Sanford Underground Research
  Facility}},
  \href{http://dx.doi.org/10.1103/PhysRevLett.112.091303}{\emph{Phys. Rev.
  Lett.} {\bfseries 112} (2014) 091303},
  [\href{https://arxiv.org/abs/1310.8214}{{\ttfamily 1310.8214}}].

\bibitem{Akerib:2015rjg}
{\scshape LUX} collaboration, D.~S. Akerib et~al., \emph{{Improved Limits on
  Scattering of Weakly Interacting Massive Particles from Reanalysis of 2013
  LUX Data}},
  \href{http://dx.doi.org/10.1103/PhysRevLett.116.161301}{\emph{Phys. Rev.
  Lett.} {\bfseries 116} (2016) 161301},
  [\href{https://arxiv.org/abs/1512.03506}{{\ttfamily 1512.03506}}].

\bibitem{Akerib:2016lao}
{\scshape LUX} collaboration, D.~S. Akerib et~al., \emph{{Results on the
  Spin-Dependent Scattering of Weakly Interacting Massive Particles on Nucleons
  from the Run 3 Data of the LUX Experiment}},
  \href{http://dx.doi.org/10.1103/PhysRevLett.116.161302}{\emph{Phys. Rev.
  Lett.} {\bfseries 116} (2016) 161302},
  [\href{https://arxiv.org/abs/1602.03489}{{\ttfamily 1602.03489}}].

\bibitem{Akerib:2016vxi}
{\scshape LUX} collaboration, D.~S. Akerib et~al., \emph{{Results from a search
  for dark matter in the complete LUX exposure}},
  \href{http://dx.doi.org/10.1103/PhysRevLett.118.021303}{\emph{Phys. Rev.
  Lett.} {\bfseries 118} (2017) 021303},
  [\href{https://arxiv.org/abs/1608.07648}{{\ttfamily 1608.07648}}].

\bibitem{Akerib:2017uem}
{\scshape LUX} collaboration, D.~S. Akerib et~al., \emph{{First searches for
  axions and axion-like particles with the LUX experiment}},
  \href{https://arxiv.org/abs/1704.02297}{{\ttfamily 1704.02297}}.

\bibitem{Akerib:2017kat}
{\scshape LUX} collaboration, D.~S. Akerib et~al., \emph{{Limits on
  spin-dependent WIMP-nucleon cross section obtained from the complete LUX
  exposure}},
  \href{http://dx.doi.org/10.1103/PhysRevLett.118.251302}{\emph{Phys. Rev.
  Lett.} {\bfseries 118} (2017) 251302},
  [\href{https://arxiv.org/abs/1705.03380}{{\ttfamily 1705.03380}}].

\bibitem{Akerib:2012ys}
{\scshape LUX} collaboration, D.~S. Akerib et~al., \emph{{The Large Underground
  Xenon (LUX) Experiment}},
  \href{http://dx.doi.org/10.1016/j.nima.2012.11.135}{\emph{Nucl. Instrum.
  Meth.} {\bfseries A704} (2013) 111--126},
  [\href{https://arxiv.org/abs/1211.3788}{{\ttfamily 1211.3788}}].

\bibitem{LHCb}
M.~Anelli, P.~Campana, E.~Dane, C.~Forti, G.~Martellotti, G.~Penso et~al.,
  \emph{Quality tests of the lhcb muon chambers at the lnf production site},
  \href{http://dx.doi.org/10.1109/TNS.2006.869836}{\emph{IEEE Transactions on
  Nuclear Science} {\bfseries 53} (Feb, 2006) 330--335}.

\bibitem{ARNISON1990431}
G.~Arnison et~al., \emph{Production and testing of limited streamer tubes for
  the end-cap muon subdetector of opal},
  \href{http://dx.doi.org/http://dx.doi.org/10.1016/0168-9002(90)90283-C}{\emph{Nuclear
  Instruments and Methods in Physics Research Section A: Accelerators,
  Spectrometers, Detectors and Associated Equipment} {\bfseries 294} (1990) 431
  -- 438}.

\bibitem{DEWULF1988109}
J.~P. DeWulf et~al., \emph{Test results and conditioning procedure of a limited
  streamer-tube calorimeter},
  \href{http://dx.doi.org/http://dx.doi.org/10.1016/0168-9002(88)91024-8}{\emph{Nuclear
  Instruments and Methods in Physics Research Section A: Accelerators,
  Spectrometers, Detectors and Associated Equipment} {\bfseries 263} (1988) 109
  -- 113}.

\bibitem{AREFIEV198971}
A.~Arefiev et~al., \emph{Proportional chambers for the barrel hadron
  calorimeter of the l3 experiment},
  \href{http://dx.doi.org/http://dx.doi.org/10.1016/0168-9002(89)90337-9}{\emph{Nuclear
  Instruments and Methods in Physics Research Section A: Accelerators,
  Spectrometers, Detectors and Associated Equipment} {\bfseries 275} (1989) 71
  -- 80}.

\bibitem{coimbra}
F.~Neves, A.~Lindote, A.~Morozov, V.~Solovov, C.~Silva, P.~Bras et~al.,
  \emph{{Measurement of the absolute reflectance of polytetrafluoroethylene
  (PTFE) immersed in liquid xenon}},
  \href{http://dx.doi.org/10.1088/1748-0221/12/01/P01017}{\emph{JINST}
  {\bfseries 12} (2017) P01017},
  [\href{https://arxiv.org/abs/1612.07965}{{\ttfamily 1612.07965}}].

\bibitem{KANIK1996455}
I.~Kanik, \emph{Far ultraviolet emission spectrum of xenon induced by electron
  impact at 100 ev},
  \href{http://dx.doi.org/http://dx.doi.org/10.1016/0009-2614(96)00688-4}{\emph{Chemical
  Physics Letters} {\bfseries 258} (1996) 455 -- 459}.

\bibitem{Policarpo}
A.~J. P.~L. Policarpo, \emph{Light production and gaseous detectors},
  {\emph{Physica Scripta} {\bfseries 23} (1981) 539}.

\bibitem{ANDRESEN1977371}
R.~D. Andresen, E.-A. Leimann and A.~Peacock, \emph{The nature of the light
  produced inside a gas scintillation proportional counter},
  \href{http://dx.doi.org/http://dx.doi.org/10.1016/0029-554X(77)90306-8}{\emph{Nuclear
  Instruments and Methods} {\bfseries 140} (1977) 371 -- 374}.

\bibitem{Munsoo}
M.~Yun, K.~Yoshino, Y.~Inuishi and M.~Kawatsu, \emph{Photoconduction in
  polytetrafluoroethylene induced by vacuum-ultraviolet light}, {\emph{Japanese
  Journal of Applied Physics} {\bfseries 21} (1982) 1592}.

\bibitem{seki1990electronic}
K.~Seki, H.~Tanaka, T.~Ohta, Y.~Aoki, A.~Imamura, H.~Fujimoto et~al.,
  \emph{Electronic structure of poly(tetrafluoroethylene) studied by ups, vuv
  absorption, and band calculations}, {\emph{Physica Scripta} {\bfseries 41}
  (1990) 167}.

\bibitem{Zhang20061995}
G.~J. Zhang, K.~Yang, W.~B. Zhao and Z.~Yan, \emph{On the surface trapping
  parameters of polytetrafluoroethylene block},
  \href{http://dx.doi.org/http://dx.doi.org/10.1016/j.apsusc.2006.03.082}{\emph{Applied
  Surface Science} {\bfseries 253} (2006) 1995 -- 1998}.

\bibitem{paulmier2014radiation}
R.~Hanna, T.~Paulmier, P.~Molinie, M.~Belhaj, B.~Dirassen, D.~Payan et~al.,
  \emph{Radiation induced conductivity in space dielectric materials},
  \href{http://dx.doi.org/10.1063/1.4862741}{\emph{Journal of Applied Physics}
  {\bfseries 115} (2014) 033713},
  [\href{https://arxiv.org/abs/http://dx.doi.org/10.1063/1.4862741}{{\ttfamily
  http://dx.doi.org/10.1063/1.4862741}}].

\bibitem{green2006ptferho}
N.~W. Green, A.~R. Frederickson and J.~R. Dennison, \emph{Experimentally
  derived resistivity for dielectric samples from the crres internal discharge
  monitor}, \href{http://dx.doi.org/10.1109/TPS.2006.883372}{\emph{IEEE
  Transactions on Plasma Science} {\bfseries 34} (Oct, 2006) 1973--1978}.

\bibitem{mellinger2004charge}
A.~Mellinger, \emph{Charge storage in electret polymers: mechanisms,
  characterization and applications}.
\newblock PhD thesis, University Potsdam, 2004.

\bibitem{zhang1991constant}
X.~Zhang-Fu, D.~Hai, Y.~Guo-Mao, L.~Ting-Ji and S.~Xi-Min,
  \emph{Constant-current corona charging of teflon pfa},
  \href{http://dx.doi.org/10.1109/14.68224}{\emph{IEEE Transactions on
  Electrical Insulation} {\bfseries 26} (Feb, 1991) 35--41}.

\bibitem{kressmann1996electrets}
R.~Kressmann, G.~M. Sessler and P.~Gunther, \emph{Space-charge electrets},
  \href{http://dx.doi.org/10.1109/94.544184}{\emph{IEEE Transactions on
  Dielectrics and Electrical Insulation} {\bfseries 3} (Oct, 1996) 607--623}.

\bibitem{comsolRef}
``{\textsc{comsol} Multiphysics\textsuperscript{\textregistered}}.''

\bibitem{mcdonald2003notes}
K.~T. McDonald, \emph{Notes on electrostatic wire grids}, {\emph{Internal note,
  available at \url{http://www.hep.princeton.edu/~mcdonald/examples/grids.pdf}}
  (2003) }.

\bibitem{Albert:2016bhh}
{\scshape EXO-200} collaboration, J.~B. Albert et~al., \emph{{Measurement of
  the Drift Velocity and Transverse Diffusion of Electrons in Liquid Xenon with
  the EXO-200 Detector}},
  \href{http://dx.doi.org/10.1103/PhysRevC.95.025502}{\emph{Phys. Rev.}
  {\bfseries C95} (2017) 025502},
  [\href{https://arxiv.org/abs/1609.04467}{{\ttfamily 1609.04467}}].

\bibitem{scott_krm}
{\scshape LUX} collaboration, D.~S. Akerib et~al., \emph{{$^{83\textrm{m}}$Kr
  calibration of the 2013 LUX dark matter search}},
  \href{https://arxiv.org/abs/1708.02566}{{\ttfamily 1708.02566}}.

\bibitem{chib1995understanding}
S.~Chib and E.~Greenberg, \emph{Understanding the metropolis-hastings
  algorithm}, {\emph{The American Statistician} {\bfseries 49} (1995)
  327--335}.

\bibitem{Akerib201263}
{\scshape LUX} collaboration, D.~Akerib et~al., \emph{Luxsim: A
  component-centric approach to low-background simulations},
  \href{http://dx.doi.org/http://doi.org/10.1016/j.nima.2012.02.010}{\emph{Nuclear
  Instruments and Methods in Physics Research Section A: Accelerators,
  Spectrometers, Detectors and Associated Equipment} {\bfseries 675} (2012) 63
  -- 77}.

\bibitem{bg_paper}
{\scshape LUX} collaboration, D.~S. Akerib et~al., ``{Backgrounds in the
  extended LUX rare event search}.''.

\bibitem{mei2012direct}
Y.~Mei, \emph{Direct Dark Matter Search with the XENON100 Experiment}.
\newblock PhD thesis, Rice University, 2012.

\bibitem{Szydagis:2013sih}
M.~Szydagis, A.~Fyhrie, D.~Thorngren and M.~Tripathi, \emph{{Enhancement of
  NEST Capabilities for Simulating Low-Energy Recoils in Liquid Xenon}},
  \href{http://dx.doi.org/10.1088/1748-0221/8/10/C10003}{\emph{JINST}
  {\bfseries 8} (2013) C10003},
  [\href{https://arxiv.org/abs/1307.6601}{{\ttfamily 1307.6601}}].

\bibitem{Aprile:2006kx}
E.~Aprile, C.~E. Dahl, L.~DeViveiros, R.~Gaitskell, K.~L. Giboni, J.~Kwong
  et~al., \emph{{Simultaneous measurement of ionization and scintillation from
  nuclear recoils in liquid xenon as target for a dark matter experiment}},
  \href{http://dx.doi.org/10.1103/PhysRevLett.97.081302}{\emph{Phys. Rev.
  Lett.} {\bfseries 97} (2006) 081302},
  [\href{https://arxiv.org/abs/astro-ph/0601552}{{\ttfamily
  astro-ph/0601552}}].

\bibitem{Lin:2015jta}
Q.~Lin, J.~Fei, F.~Gao, J.~Hu, Y.~Wei, X.~Xiao et~al., \emph{{Scintillation and
  ionization responses of liquid xenon to low energy electronic and nuclear
  recoils at drift fields from 236 V/cm to 3.93 kV/cm}},
  \href{http://dx.doi.org/10.1103/PhysRevD.92.032005}{\emph{Phys. Rev.}
  {\bfseries D92} (2015) 032005},
  [\href{https://arxiv.org/abs/1505.00517}{{\ttfamily 1505.00517}}].

\bibitem{Baudis:2013cca}
L.~Baudis, H.~Dujmovic, C.~Geis, A.~James, A.~Kish, A.~Manalaysay et~al.,
  \emph{{Response of liquid xenon to Compton electrons down to 1.5 keV}},
  \href{http://dx.doi.org/10.1103/PhysRevD.87.115015}{\emph{Phys. Rev.}
  {\bfseries D87} (2013) 115015},
  [\href{https://arxiv.org/abs/1303.6891}{{\ttfamily 1303.6891}}].

\bibitem{Chan:1995tk}
A.~Chan et~al., \emph{{Performance of the HPC calorimeter in DELPHI}},
  \href{http://dx.doi.org/10.1109/23.467923}{\emph{IEEE Trans. Nucl. Sci.}
  {\bfseries 42} (1995) 491--498}.

\bibitem{Akerib:2015wdi}
{\scshape LUX} collaboration, D.~S. Akerib et~al., \emph{{Tritium calibration
  of the LUX dark matter experiment}},
  \href{http://dx.doi.org/10.1103/PhysRevD.93.072009}{\emph{Phys. Rev.}
  {\bfseries D93} (2016) 072009},
  [\href{https://arxiv.org/abs/1512.03133}{{\ttfamily 1512.03133}}].

\bibitem{Akerib:2016mzi}
{\scshape LUX} collaboration, D.~S. Akerib et~al., \emph{{Low-energy (0.7-74
  keV) nuclear recoil calibration of the LUX dark matter experiment using D-D
  neutron scattering kinematics}},
  \href{https://arxiv.org/abs/1608.05381}{{\ttfamily 1608.05381}}.

\end{thebibliography}\endgroup
%\include{run4_fields}
%\end{thebibliography}

\end{document}